\documentclass[%
reprint,
amsmath, amssymb,
aps,
prd
]{revtex4-2}

\usepackage[colorlinks = true, linkcolor = blue]{hyperref}
\usepackage{xcolor}
\usepackage{ulem}
\usepackage{graphicx}
\graphicspath{{./figures/}}
\usepackage{makecell}
\usepackage{subcaption}
\usepackage{multirow}

\usepackage{float}
\usepackage{overpic}
\usepackage{comment}
\usepackage{cancel}

\usepackage{siunitx}

\usepackage{amsmath}
\usepackage[e]{esvect}
\usepackage[capitalise]{cleveref} 
\usepackage{amssymb}
\usepackage{slashed}
\usepackage{physics}

\usepackage{dcolumn}
\usepackage{bm}
\hypersetup{colorlinks, linkcolor = [rgb]{0, 0, 0.5}, citecolor = [rgb]{0,0.0,0.5}, urlcolor = [rgb]{0,0.0,0.5}}

\usepackage{pifont} 
\usepackage{tikz}

\begin{document}

\title{Analysis of Nuclear Fragmentation Functions for Pions with $A$ and $\nu$ Dependence}

\author{Mengyang Li$^{1}$, Zijian Ye$^{2,3}$, Jun Gao$^{1,4}$, XiaoMin~Shen$^{3,5}$, Hongxi Xing$^{2,4,6}$, Yuxiang Zhao$^{3,4,5,7}$}

\affiliation{
    $^1${State Key Laboratory of Dark Matter Physics, Shanghai Key Laboratory for Particle Physics and Cosmology, Key Laboratory for Particle Astrophysics and Cosmology (MOE), School of Physics and Astronomy, Shanghai Jiao Tong University, Shanghai 200240, China}
    \\
    $^2${State Key Laboratory of Nuclear Physics and Technology, Institute of Quantum Matter, South China Normal University, Guangzhou 510006, China}
    \\
    $^3${Institute of Modern Physics, Chinese Academy of Sciences, Lanzhou, Gansu 730000, China}
    \\
    $^4${Southern Center for Nuclear-Science Theory (SCNT), Institute of Modern Physics, Chinese Academy of Sciences, Huizhou 516000, China}
    \\
    $^5${University of Chinese Academy of Sciences, Beijing 100049, China}
    \\
    $^6${Guangdong Basic Research Center of Excellence for Structure and Fundamental Interactions of Matter, Guangdong Provincial Key Laboratory of Nuclear Science, Guangzhou 510006, China}
    \\
    $^7${Institute of Particle Physics, Central China Normal University, Wuhan 480079, China}
}

\email[Electronic address: ]{\\
{limengyang@sjtu.edu.cn} \\
{yezj@m.scnu.edu.cn} \\
{jung49@sjtu.edu.cn} \\
{xiaominshen@impcas.ac.cn} \\
{hxing@m.scnu.edu.cn} \\
{yxzhao@impcas.ac.cn} \\
}

\begin{abstract}

We present a QCD analysis of pion nuclear fragmentation functions (nFFs), which encode nuclear modifications to hadronization in high-energy nuclear collisions. 
Within this framework, vacuum fragmentation functions and their nuclear modifications are extracted simultaneously.
The nuclear effects are parameterized as functions of the mass number $A$, the energy of the fragmenting parton in the target rest frame $\nu$, and the hadron energy fraction $z$, allowing their dependence on these variables to be quantified.
Our analysis includes semi-inclusive deep-inelastic scattering data on nuclear targets, with kinematic cuts chosen to ensure the applicability of perturbative QCD and collinear factorization. 
The resulting fit provides a good description of most datasets, with the nFFs well constrained in the energy fraction range $z \in [0.2,\ 0.7]$. 
Additionally, with our new nFFs, we present next-to-leading order predictions in $pp$ and $pA$ collisions, which show reasonable agreement with ALICE data within the current experimental uncertainties.

\end{abstract}

\maketitle

\pagebreak
\newpage

\section{Introduction}

Understanding hadronization dynamics is essential for exploring color confinement in QCD. Fragmentation functions (FFs)~\cite{Field:1977fa,Metz:2016swz} provide the standard nonperturbative framework for describing hadron production in the final state, encoding the transition of partons from hard scattering processes into observed hadrons~\cite{Collins:1989gx,Collins:1981uk,Collins:1981uw} in single-inclusive $e^+e^-$ annihilation (SIA), semi-inclusive deep-inelastic scattering (SIDIS), and proton-proton ($pp$) collisions. Similar to parton distribution functions (PDFs)~\cite{Forte:2010dt,Forte:2013wc,Rojo:2015acz,Kovarik:2019xvh}, these functions have been proven to be universal and cannot be computed perturbatively. A reliable way is to extract various FFs from world data through global QCD analysis.

While PDFs and FFs provide a robust framework for factorized cross-section calculations in the vacuum, their application to nuclear environments requires additional considerations. 
When hard scattering processes occur in nuclei rather than free nucleons, the initial-state parton distributions are modified and described by nuclear parton distribution functions (nPDFs)~\cite{deFlorian:2003qf,Hirai:2007sx,Eskola:2010jh,Klasen:2023uqj, Duwentaster:2022kpv, Derakhshanian:2026zkx, Klasen:2025ekj, Klasen:2024xqn, Kotikov:2025qbh}. These nPDFs successfully describe inclusive observables such as nuclear deep-inelastic scattering (DIS) and Drell-Yan production. However, they alone cannot account for the observed hadron-species-dependent differences in production processes when comparing heavy nuclear targets with light nuclei or proton ($p$) targets~\cite{E665:1993mkg, HERMES:2009uge, PHENIX:2006mhb, STAR:2003oii, STAR:2006xud, Grebenyuk:2007pja}.

\begin{table*}[htb]
\centering
\small
\setlength{\tabcolsep}{6pt}
\begin{tabular}{cccccccccc}

\# & Exp & year & $\sqrt{s}$ [GeV] & Target(A) & Particle & binning & obs & Corr. source(\%) & $N_{pt}$ \\
\hline\hline

1 & HERMES~\cite{HERMES:2001ghm} & 2001 & 27.5 & $p$ & $\pi^0$ & $z$
& $\frac{1}{N_{DIS}}\frac{dN^{\pi^0}}{dz}$ & 9 & 10 \\[1mm]

2 & HERMES~\cite{HERMES:2007plz} & 2007 & 27.6 & $He$ & $\pi^{\pm}$ & $z,\nu$
& $R_A^h$ & 3 & 16 \\[1mm]

3 & & & & $He$ & $\pi^{\pm}$ & $z,Q^2$
& $R_A^h$ & 3 & 20 \\[1mm]

4 & & & & $He,Ne,Kr,Xe$ & $\pi^+$ & $z$
& $R_A^h$ & 3 & 24 \\[1mm]

5 & & & & $He,Ne,Kr,Xe$ & $\pi^-$ & $z$
& $R_A^h$ & 3 & 24 \\[1mm]

6 & & & & $He,Ne,Kr,Xe$ & $\pi^0$ & $z$
& $R_A^h$ & 3 & 24 \\[1mm]

7 & HERMES~\cite{HERMES:2011qjb} & 2011 & 27.6 & $Ne,Kr,Xe$ & $\pi^+$ & $z,Q^2$
& $R_A^h$ & 3 & 60 \\[1mm]

8 & & & & $Ne,Kr,Xe$ & $\pi^-$ & $z,Q^2$
& $R_A^h$ & 3 & 60 \\[1mm]

9 & & & & $Ne,Kr,Xe$ & $\pi^+$ & $z,\nu$
& $R_A^h$ & 3 & 51 \\[1mm]

10 & & & & $Ne,Kr,Xe$ & $\pi^-$ & $z,\nu$
& $R_A^h$ & 3 & 51 \\[1mm]

11 & HERMES~\cite{HERMES:2012uyd} & 2013 & 27.6 & $p$ & $\pi^+$ & $z,Q^2$
& $\frac{1}{N_{DIS}}\frac{dN^{\pi^+}}{dzdQ^2}$ & 0 & 25 \\[1mm]

12 & & & & $p$ & $\pi^-$ & $z,Q^2$
& $\frac{1}{N_{DIS}}\frac{dN^{\pi^-}}{dzdQ^2}$ & 0 & 25 \\[1mm]

13 & & & & $D$ & $\pi^+$ & $z,Q^2$
& $\frac{1}{N_{DIS}}\frac{dN^{\pi^+}}{dzdQ^2}$ & 0 & 25 \\[1mm]

14 & & & & $D$ & $\pi^-$ & $z,Q^2$
& $\frac{1}{N_{DIS}}\frac{dN^{\pi^-}}{dzdQ^2}$ & 0 & 25 \\[1mm]

15 & CLAS~\cite{CLAS:2021jhm} & 2021 & 5.0 & $C,Fe,Pb$ & $\pi^+$ & $z,Q^2,\nu$
& $R_A^h$ & 3 & 117 \\[1mm]

16 & & & & $C,Fe,Pb$ & $\pi^-$ & $z,Q^2,\nu$
& $R_A^h$ & 3 & 114 \\[1mm]

\multicolumn{9}{r}{\textbf{SIDIS Total:}} & 671 \\
\cline{10-10}
\hline

17 & SLD~\cite{SLD:2003ogn} & 2003 & 91.2 & SIA c-tagged & $\pi^{\pm}$ & $z$
& $\frac{1}{N_{\rm evt}}\frac{dN^{\pi^{\pm}}}{dz}$ & 1 & 21 \\[1mm]

18 & & & &SIA b-tagged & $\pi^{\pm}$ & $z$
& $\frac{1}{N_{\rm evt}}\frac{dN^{\pi^{\pm}}}{dz}$ & 1 & 21 \\[1mm]

\multicolumn{9}{r}{\textbf{SIA Total:}} & 42 \\
\cline{10-10}
\hline
\multicolumn{9}{r}{\textbf{Total:}} & 713 \\

\end{tabular}
\caption{Summary of experimental datasets included in the analysis, together with the corresponding correlated uncertainties.}
\label{t.ex}
\end{table*}

Even though such differences can be attributed to a variety of conceivable mechanisms beyond initial-state modifications of parton densities~\cite{Arleo:2008dn}, including, for instance, parton energy loss caused by medium-induced multiple gluon emission~\cite{Baier:2000mf,Kovner:2003zj,Gyulassy:2003mc,Majumder:2007iu}, it has been proposed that QCD factorization can be extended to incorporate final-state nuclear effects through the introduction of nuclear fragmentation functions (nFFs) with an explicit dependence on nuclear mass number $A$~\cite{Sassot:2009sh}. The idea was further explored in Ref.~\cite{Doradau:2024wli}, and has been extended to the three-dimensional imaging of PDFs and FFs in nuclei~\cite{Alrashed:2021csd,Alrashed:2023xsv,Barry:2023qqh}.

Although such approaches based on $A$-dependent nFFs provide a successful description of nuclear modifications across a broad range of data, they do not explicitly account for additional kinematic variables that may influence the hadronization process in nuclear environments. 
In particular, recent high-precision measurements from HERMES~\cite{HERMES:2007plz,HERMES:2011qjb} and CLAS~\cite{CLAS:2021jhm} suggest that the strength of nuclear modifications exhibits a nontrivial dependence on the virtual-photon energy $\nu$ in the target rest frame, indicating that the dynamics of in-medium hadronization cannot be fully captured by $A$-dependence.
This experimental observation motivates the present study. In this work, we introduce a parameterization of nuclear modifications that depends on $A$ and $\nu$, and simultaneously extract the vacuum FFs and their nuclear corrections. The implementation of the Hessian method in our study further improves the reliability of the extraction of nFFs by constructing uncertainty eigenvector sets for error quantification. 
In addition, updated nPDF sets, including nCTEQ15WZ~\cite{Kusina:2020lyz}, nNNPDF3.0~\cite{AbdulKhalek:2022fyi}, and EPPS21~\cite{Eskola:2021nhw}, are used to perform systematic comparisons and assess the associated theoretical uncertainties. Heavy-quark-tagged SIA data are also included to improve constraints on heavy-flavor fragmentation.

The paper is organized as follows. Section~\ref{s.data} describes the experimental datasets used in this analysis, including SIDIS measurements from HERMES and CLAS and heavy-flavor-tagged SIA data from SLD. Section~\ref{s.theory} presents the theoretical framework for vacuum FFs and nFFs, including details of the theoretical calculations and the definition of the goodness-of-fit function. 
The extraction of vacuum FFs and the corresponding predictions compared with data are presented in Section~\ref{s.vff}, which constrain the theoretical conditions for the subsequent determination of nFFs and define the corresponding default setup. Based on this, the extracted nFFs and their corresponding comparisons with data are discussed in Section~\ref{s.nff}. Predictions for $\pi$ production in $pp$ and proton-lead ($p$Pb) collisions at ALICE are presented in Section~\ref{s.prediction}. Finally, Section~\ref{s.conclusion} summarizes our main findings.

\section{Experimental data sets fitted}
\label{s.data}

We begin by introducing the key observables in nuclear SIDIS, i.e. $\ell+A \to \ell'+h+X$. 
The four-momenta of the target nucleon, exchanged virtual photon, incoming lepton, and produced hadron are denoted by $P$, $q$, $\ell$, and $P_h$, respectively.
The differential multiplicity of identified hadrons is defined as
\begin{equation}
\frac{\mathrm{d}M^{h}(x,z,Q^{2})}{\mathrm{d}z} = \frac{\mathrm{d}^{3}\sigma^{h}(x,z,Q^{2})/\mathrm{d}x\,\mathrm{d}Q^{2}\,\mathrm{d}z}{\mathrm{d}^{2}\sigma^{\mathrm{DIS}}(x,Q^{2})/\mathrm{d}x\,\mathrm{d}Q^{2}},
\label{eq.multi}
\end{equation}
where $x = Q^{2}/(2P \cdot q)$ is the Bjorken variable, $z = (P \cdot P_h)/(P \cdot q)$ is the hadron energy fraction, and $Q^{2} = -q^{2}$ denotes the virtuality of the exchanged photon. In this process, we expect the nuclear modification from initial-state nPDFs to be largely canceled, therefore providing a golden channel for the global extraction of nFFs.

To quantify nuclear modifications, the multiplicity ratio $R_A^h$ is introduced as the ratio of the differential hadron multiplicity for a nuclear target with mass number $A$ to that for a deuterium (D) target.
Since the virtual-photon energy $\nu$ in the target rest frame provides an additional kinematic variable in nuclear SIDIS, we keep it explicitly in the nuclear multiplicity ratio in order to test possible $\nu$-dependent effects. 
The multiplicity ratio is then defined as
\begin{equation}
R_A^h(x,z,Q^2,\nu)
=
\frac{
\mathrm{d}M_A^h(x,z,Q^2,\nu)/\mathrm{d}z
}{
\mathrm{d}M_D^h(x,z,Q^2,\nu)/\mathrm{d}z
},
\label{e.SIDIS-Rh}
\end{equation}
where $\mathrm{d}M_A^h/\mathrm{d}z$ and $\mathrm{d}M_D^h/\mathrm{d}z$ denote the corresponding differential multiplicities for the nuclear target with mass number $A$ and the D target, respectively.

In Table~\ref{t.ex}, we present a comprehensive compilation of the experimental data used in this analysis, including SIDIS measurements from HERMES~\cite{HERMES:2001ghm,HERMES:2007plz,HERMES:2011qjb,HERMES:2012uyd} and CLAS~\cite{CLAS:2021jhm} and heavy-flavor-tagged SIA measurements from SLD~\cite{SLD:2003ogn}.  
The HERMES data span multiple years and include measurements on $p$, helium (He), neon (Ne), krypton (Kr), xenon (Xe), and D targets, with final-state hadrons $\pi^+$, $\pi^-$, and $\pi^0$. These data are presented in various kinematic bins of $z$, $Q^2$, and $\nu$, and are reported either as the double ratio $R_A^h$ or as the normalized differential multiplicity $dN_h / (dz dQ^2)$. The data cover a wide range of nuclear targets, providing valuable constraints on the $A$-dependence of nFFs. It should be noted that the HERMES data reported in 2011~\cite{HERMES:2011qjb} are not statistically independent of earlier HERMES measurements~\cite{HERMES:2007plz}. The 2011 HERMES results present nuclear multiplicity ratios in two-dimensional kinematic bins, while the earlier results were mostly shown as functions of a single kinematic variable. Therefore, including the two HERMES measurements in the fit simultaneously can introduce a certain degree of double counting. Nevertheless, in order to make maximal use of the available experimental information, we include both measurements in the present analysis. We note that several statistically correlated HERMES data sets have also been fitted simultaneously in global QCD analyses~\cite{Doradau:2024wli}.
More recently, the CLAS collaboration has published high-precision measurements on carbon (C), iron (Fe), and lead (Pb) targets, covering a broad range of $z$, $Q^2$, and $\nu$~\cite{CLAS:2021jhm}. These data extend the nuclear coverage to heavier nuclei and are reported as $R_A^h$ in multiple $\nu$ bins, enabling a detailed study of nuclear medium effects across different energy regimes. The combined dataset, as summarized in Table~\ref{t.ex}, offers a robust and diverse set of observables for constraining nFFs.
To further constrain the vacuum FFs, we also include SLD SIA data~\cite{SLD:2003ogn} on charm- and bottom-tagged events, which provide direct constraints on the heavy flavor FFs.

We apply kinematic cuts to ensure the reliability of the QCD collinear factorization framework. For both SIA and SIDIS data sets, we require $z>0.1$. To further suppress regions where power corrections may become important, we impose an additional requirement on the hadron energy $E_h$, which is evaluated in the center-of-mass frame for SIA and in the Breit frame for SIDIS, by requiring the corresponding hadron energy scale, $zQ/2$, to satisfy $zQ/2>0.19~\mathrm{GeV}$.


\section{Framework for Medium-Modified Fragmentation Functions}
\label{s.theory}
\subsection{QCD Factorization and vacuum FFs}
\label{ss.vacuum}

In the collinear factorization framework, the cross sections can be factorized into perturbatively calculable short-distance partonic cross sections, and non-perturbative distribution functions \cite{Collins:1989gx}. 
For example, the SIDIS differential cross section can be written as
\begin{equation}
\begin{aligned}
\frac{d^{3}\sigma^{\ell p \rightarrow \ell h X}}{dx\,dy\,dz}
&=
\frac{2\pi \alpha_{\rm em}^{2}}{Q^{2}}
\Bigg[
\frac{1+(1-y)^{2}}{y}\, F_{T}^{h}(x,z,Q^{2})
\\
&\quad +
\frac{2(1-y)}{y}\, F_{L}^{h}(x,z,Q^{2})
\Bigg],
\end{aligned}
\end{equation}
where $F_{T}^{h}$ and $F_{L}^{h}$ are the semi-inclusive structure functions, and $y = (P \cdot q)/(P \cdot \ell)$ is the inelasticity.

At next-to-leading order (NLO), the SIDIS structure functions can be expressed as:
\begin{equation}
\begin{aligned}
\label{e.structure_1}
F_T^h(x,z,Q^2) &= \sum_q e_q^2 \Bigg[
f_1^{q/p}(x,Q^2)\,D_q^h(z,Q^2)
\\
&\quad + \frac{\alpha_s(Q^2)}{2\pi}
\biggl(
f_1^{q/p}\otimes C_1^{qq}\otimes D_q^h
\\
&\quad + f_1^{q/p}\otimes C_1^{gq}\otimes D_g^h
\\
&\quad + f_1^{g/p}\otimes C_1^{qg}\otimes D_q^h
\biggr)
\Bigg],
\end{aligned}
\end{equation}
\begin{equation}
\begin{aligned}
\label{e.structure_2}
F_L^h(x,z,Q^2) &= \frac{\alpha_s(Q^2)}{2\pi}\sum_q e_q^2 \Bigg[
f_1^{q/p}\otimes C_L^{qq}\otimes D_q^h
\\
&\quad + f_1^{q/p}\otimes C_L^{gq}\otimes D_g^h
\\
&\quad + f_1^{g/p}\otimes C_L^{qg}\otimes D_q^h
\Bigg],
\end{aligned}
\end{equation}
where $f_1^{i/p}$ denotes the PDFs, while FFs ($D_i^h$) describe the non-perturbative hadronization of a final-state parton of flavor $i$ into an observed hadron $h$. The functions $C_{1(L)}^{ij}$ are the NLO perturbatively calculable SIDIS coefficient functions~\cite{Nason:1993xx, Furmanski:1981cw, Graudenz:1994dq,deFlorian:1997zj,deFlorian:2012wk}, and the NNLO coefficient functions can be found in \cite{Goyal:2023zdi,Bonino:2024qbh}. Approximate NNLO and N\textsuperscript{3}LO (next-to-next-to-next-to-leading order) structure functions have also been obtained through expansions of threshold resummation expressions~\cite{Abele:2021nyo,Abele:2022wuy,He:2025hin,Dong:2026uvw}. The symbol $\otimes$ denotes the standard convolution integral defined as 
\begin{equation}
\begin{aligned}
    [f\otimes g](z)=\int_0^1dx\int_0^1dy f(x)g(y)\delta(z-xy).
\end{aligned}
\end{equation}

Focusing on the FFs, $D_i^h$ are universal within the framework of QCD factorization in the vacuum and cannot be computed perturbatively, and can be extracted from world data through global QCD analysis, similar to PDFs~\cite{Forte:2010dt,Forte:2013wc, Rojo:2015acz, Kovarik:2019xvh}. Such analyses include hadron production in SIA, SIDIS, and $pp$ collisions, with energy scale satisfying $Q^2 \gg \Lambda_{\rm QCD}^2$.
The evolution of FFs $D^h_i$ with respect to the scale $Q^2$ is described by the Dokshitzer-Gribov-Lipatov-Altarelli-Parisi (DGLAP) equation
\cite{Gribov:1972ri,Lipatov:1974qm,Altarelli:1977zs,Dokshitzer:1977sg}:
\begin{equation}
\label{e.DGLAP}
\begin{aligned}
    \frac{\partial}{\partial {\rm ln} Q^2}D_i^h(z, Q^2)=\sum_j [P_{ji} \otimes D_j^h] (z,Q^2),
\end{aligned}
\end{equation}
where $P_{ji}$ are the time-like splitting functions, describing the probability for a parton $i$ to split into a parton $j$. These time-like splitting functions are currently known up to $\mathcal{O}(\alpha_s^3)$~\cite{Mitov:2006ic,Moch:2007tx,Almasy:2011eq,Chen:2020uvt,Ebert:2020qef,Luo:2020epw}, providing the theoretical basis for high-precision global analyses of FFs. It should be noted that, although time-like and space-like splitting functions (the latter relevant for PDFs) are identical at leading order, they differ beyond NLO~\cite{Curci:1980uw,Furmanski:1980cm}.

\subsection{Nuclear FFs and Parametrization}

\label{ss.nuclear}
Assuming that the QCD collinear factorization for inclusive production is valid also for collisions involving nuclei, one can naturally define fragmentation functions $D_{i/A}^h(z,Q^2)$ for nuclear environment characterized by mass number $A$~\cite{Sassot:2009sh,Doradau:2024wli}. In these factorization approaches, $D_{i/A}^h(z,Q^2)$ parametrize nuclear modifications solely through their dependence on $A$ as:
\begin{equation}
D_{i/A}^h(z,Q_0^2)=\int_z^1\frac{dy}{y}W_i^h(y,A,Q^2_0)D_i^h(\frac{z}{y},Q_0^2),
\label{eq:base_model}
\end{equation}
where the subscript $i$ labels different parton flavors, the superscript $h$ labels different hadron species, and $W_i^h(y,A,Q^2_0)$ encodes the nuclear modification dependence derived from fixed-target hadron-nucleus data~\cite{deFlorian:2003qf,Hirai:2007sx}, and has a complex parameterization. A similar reasoning is applied with great phenomenological success in analyses of nPDFs that account for medium-induced effects in the initial-state~\cite{deFlorian:2003qf,Hirai:2007sx,Eskola:2010jh}.

\begin{table}[b]
\setcellgapes{2.5 pt}
\makegapedcells
\begin{tabular}{|c|c|c|c|c|c|}
\hline
 & flavor & $a_0$ & $\alpha$ & $\beta$ & $a_1$ \\
\hline
\multirow{6}{*}{$D_i^{h}(z,Q^2)$} 
& $u = \overline{d}$ & \ding{51} & \ding{51}  & \ding{51} & \ding{51} \\
\cline{2-6}
& $d = \overline{u}$ & \ding{51} & \ding{51}  & \ding{51} & \ding{51} \\
\cline{2-6}
& $s = \overline{s}$ & $=a_{0,d}$    & $= \alpha_d$  & $= \beta_d$ & $= a_{1,d}$ \\
\cline{2-6}
& $c = \overline{c}$ & \ding{51} & \ding{51}  & \ding{51} & - \\
\cline{2-6}
& $b = \overline{b}$ & \ding{51} & \ding{51}  & \ding{51} & - \\
\cline{2-6}
& $g$                & \ding{51} & \ding{51}  & \ding{51} & - \\
\hline

\multirow{3}{*}{$D_i^{\prime h}(z,Q^2)$} 
& $u = \overline{d}$  & \ding{51} & \ding{51}  & \ding{51} & \ding{51} \\
\cline{2-6}
& $d = \overline{u}$  & \ding{51} & \ding{51}  & \ding{51} & \ding{51} \\
\cline{2-6}
& $s = \overline{s}$ & $=a_{0,m,d}$    & $= \alpha_{m,d}$  & $= \beta_{m,d}$ & $= a_{1,m,d}$  \\
\hline
\end{tabular}
\caption{
Summary of the parameterization for $\pi^+$ vacuum FFs and nuclear modification terms.
Check marks under each parameter mean that the parameter is free to vary.
}
\label{t.parameterization}
\end{table}

However, such models may not fully capture the possible dependence of nuclear modifications on the virtual-photon energy $\nu$ in the target rest frame, which is directly accessible in $eA$ measurements. Recent precision measurements~\cite{HERMES:2011qjb,CLAS:2021jhm} provide data over a broad range of $\nu$, offering an opportunity to test whether additional energy dependence is favored by the data. Such a dependence may be connected to the formation time of hadrons. At larger $\nu$, the struck parton carries more energy, which may modify the space-time development of hadronization and thereby change the strength of final-state interactions in the nuclear medium.

Motivated by this possibility, we introduce a phenomenological dependence on both $A$ and $\nu$ into the nFF framework, and examine whether the available SIDIS data prefer such an extension over an $A$-only description. Within the collinear factorization framework for nuclear systems, we propose a model that decomposes the nFFs $D_{i/A}^h(z,Q^2,\nu)$ into vacuum FFs $D_i^{h}(z,Q^2)$ and a modified term $D_i^{\prime h}(z,Q^2)$:

\begin{equation}
\begin{aligned}
D_{i/A}^h(z,Q^2,\nu) &= D_i^h(z,Q^2) -  \mathcal{F}_A(A,\nu) \times D_i^{\prime h}(z,Q^2), \\
\end{aligned}
\label{eq:nFF_modification}
\end{equation}
where the $D_i^{h}(z,Q^2)$, $D_i^{\prime h}(z,Q^2)$ and $\mathcal{F}_A$ are constrained simultaneously in this work.
In this parametrization, $D_i^{\prime h}(z,Q^2)$ can be either positive or negative in principle, allowing the model to describe both suppression and enhancement effects. The minus sign convention is chosen because our analysis focuses on the medium to large $z$ region where experimental data predominantly show suppression effects. Consequently, the fitted $D_i^{\prime h}(z,Q^2)$ is expected to be positive.

To parameterize the $D_{i/A}^h$ at the initial scale $Q_0$, we adopt the following functional forms:
\begin{equation}
\begin{aligned}
zD_i^h(z,Q_0) &= z^{\alpha_i^{h}}(1-z)^{\beta_i^{h}}\exp\left(\sum_{n=0}^k a_{i,n}^{h} z^{n/2}\right), \\
zD_i^{\prime h}(z,Q_0) &= z^{{\alpha'}_i^{h}}(1-z)^{{\beta'}_i^{h}}\exp\left(\sum_{n=0}^k {a}_{i,n}^{\prime h} z^{n/2}\right),
\end{aligned}
\label{eq.nffpar}
\end{equation}
where $\alpha,\beta,a$ are the fitted parameters. By varying $k=0,1,2\cdots$, one can freely choose different parameterizations. In practice, we increase the value of $k$ until no discernible improvement in fit quality can be obtained. In this framework, we assume that the factorization of the cross section used in vacuum in Eq.~\eqref{e.structure_1} and Eq.~\eqref{e.structure_2} remains applicable to nuclear targets, with the vacuum FFs replaced by the nFFs. Accordingly, both $D_i^{h}(z,Q^2)$ and $D_i^{\prime h}(z,Q^2)$ should obey the DGLAP evolution equations.

The parameterization for $\pi$ is summarized in Table~\ref{t.parameterization}. By assuming charge-conjugation symmetry, nFFs for $\pi^-$ can be obtained from $\pi^+$. To reduce the number of free parameters, we impose flavor symmetries between favored and unfavored light (anti)quark FFs, as indicated by the equality signs in the table, where favored $u$ and $\bar{d}$ quarks share the same values of $\alpha$, $\beta$, $a_0$, and $a_1$ at the initial scale. Identical distributions are also imposed among the unfavored (anti)quarks $\bar{u}$, $d$, $s$, and $\bar{s}$. Nuclear modifications are assumed to apply only to light quarks, with similar flavor symmetries adopted in light quark contributions.

Finally, $\mathcal{F}_A(A,\nu)$ encodes the nuclear modifications, incorporating both $A$ and $\nu$ dependence, via:
\begin{equation}
\begin{aligned}
\mathcal{F}_A(A,\nu)=\frac{f(A)}{(\nu/\nu_0)^{\delta}}. \\
\end{aligned}
\label{eq.fa}
\end{equation}
Here, $\nu_0=1~\mathrm{GeV}$ is introduced as a reference energy scale to ensure that $\mathcal{F}_A(A,\nu)$ is dimensionless. The function $f(A)$ parametrizes the nuclear dependence, with an independent parameter assigned to each nucleus, and will be discussed in Section~\ref{ss.nff_modification}. The exponent $\delta$ controls the strength of the $\nu$ dependence. In this work, we consider three representative choices, $\delta=0$, $0.5$, and $1$. The case $\delta=0$ corresponds to an $A$-dependent but $\nu$-independent modification, and is included as a reference scenario. For $\delta>0$, the modification factor $\mathcal{F}_A(A,\nu)$ decreases with increasing $\nu$ and vanishes in the limit $\nu\to\infty$, so that $D_{i/A}^h$ approaches the vacuum FF $D_i^h$.

\subsection{Theoretical computations}
\label{ss.theoretical}

The theoretical computation techniques have been summarized in the previous NPC23 studies~\cite{Gao:2024dbv, Gao:2025bko, Gao:2025hlm}. Here we briefly recall the main ingredients.
In this work, the main analysis is carried out at NLO accuracy, while NNLO results are used in Section~\ref{s.vff} as a benchmark to assess higher-order effects and perturbative stability. 
For the NLO analysis, the FFs are evolved using the two-loop time-like splitting kernels, which were calculated in Refs.~\cite{Stratmann_1997}. 
For the NNLO comparison, the evolution is performed with the three-loop time-like splitting kernels~\cite{Mitov:2006ic,Moch:2007tx,Almasy:2011eq,Chen:2020uvt}. The corresponding NLO and NNLO DGLAP evolutions are carried out with HOPPET~\cite{Salam_2009,salam2008hoppet}.
Theoretical calculations of differential cross sections are carried out up to NNLO in QCD using the FMNLO program~\cite{Gao:2024dbv,Liu:2023fsq}, which can generate and store interpolation tables of the coefficient functions, ensuring fast convolution with arbitrary FFs without repeating the calculations. 
Furthermore, in this work, the dependence on $\nu$ is taken into account, and the corresponding interpolation grid construction is extended accordingly to incorporate this dependence, ensuring consistency in the nFFs analysis.

We adopt a zero-mass variable flavor number scheme (ZM-VFNS), in which heavy quarks FFs are non-zero but do not evolve until the mass thresholds are reached with maximum of $n_f=5$, specifically at $m_c =1.4$ GeV and $m_b = 4.5$ GeV for charm and bottom quarks. The strong coupling constant is consistently taken as $\alpha_S(M_Z)=0.118$ throughout the calculations. For theoretical predictions of hadron production at SIA with heavy-flavor-tagged events, we only include contributions from Feynman diagrams with the specified heavy quark coupled directly to the $Z$ boson or photon, which is well justified at NLO. There are ambiguities on matching theoretical predictions to the experimental measurements when going beyond NLO, e.g., on treatment of contributions from gluon splitting into heavy quarks.

The central values for the renormalization and fragmentation scales ($\mu_{R,0}$ and $\mu_{D,0}$) are set to the momentum transfer $Q$ for both SIA and SIDIS. The factorization scale ($\mu_{F,0}$) of initial hadrons for SIDIS is also set to $Q$.

\subsection{Goodness of fit function and the covariance matrix}
\label{s.goodness}
The agreement between the data points $D_{k}$ and the corresponding theoretical predictions $T_{k}$ is quantified by the $\chi^2$ function~\cite{Pumplin_2002}:

\begin{equation}
\chi^{2}(\{a\},\{\lambda\})=\sum_{k=1}^{N_{\mathrm{pt}}}\frac{1}{s_{k}^{2}} \left(D_{k}-T_{k}-\sum_{\alpha=1}^{N_{\lambda}}\beta_{k,\alpha}\lambda_{\alpha} \right)^{2}+\sum_{\alpha=1}^{N_{\lambda}}\lambda_{\alpha}^{2},
\end{equation}
where $\{a\}$ are parameters of FFs, the nuisance parameters $\{\lambda\}$ describe sources of correlated errors, which are assumed to follow standard normal distributions, $s_{k}$ represents the total uncorrelated systematic and statistical errors, and $\beta_{k,\alpha}$ quantifies the sensitivity of the $k$-th measurement to the $\alpha$-th correlated error source. In our case, the correlated errors include the normalization uncertainties of the measurements, listed in the penultimate column of Table~\ref{t.ex}. When theoretical uncertainties are included, they are estimated from scale variations and taken as half the width of the scale-variation band, following Ref.~\cite{Liu:2023fsq}.

\begin{table*}[httb]
\centering
\small 
~\\
\setlength{\tabcolsep}{4pt} 
\begin{tabular}{cccccccccccccc}

Exp & Target(A) & Particle & binning & $N_{pt}$ &\multicolumn{4}{c}{w/ theo. unc.($\chi^2/N_{pt}$)} &  \multicolumn{4}{c}{w/o theo. unc.($\chi^2/N_{pt}$)}\\
&&&&&\multicolumn{2}{c}{$Q_0=1.0$ GeV}&\multicolumn{2}{c}{$Q_0=1.3$ GeV}&\multicolumn{2}{c}{$Q_0=1.0$  GeV}&\multicolumn{2}{c}{$Q_0=1.3$ GeV} \\
&&&&&NLO&NNLO&NLO&NNLO&NLO&NNLO&NLO&NNLO\\ 
\hline\hline
HERMES~\cite{HERMES:2001ghm}   & $p$ &$\pi^0$ &  $z$ & 10 & 0.56 & 1.48 & 0.50 & 1.59 & 2.51 & 3.18 & 2.93 & 3.92 \\[1mm]
HERMES~\cite{HERMES:2012uyd}   & $p$ &$\pi^+$ &  $z, Q^2$ & 25 & 0.88 & 1.48 & 0.64 & 1.33 & 1.16 & 3.73 & 0.87 & 2.94 \\[1mm]
& $p$ &$\pi^-$ &  $z, Q^2$ & 25 & 0.74 & 0.99 & 0.48 & 0.83 & 0.91 & 4.76 & 0.55 & 3.89 \\[1mm]
& $D$ &$\pi^+$ &  $z, Q^2$ & 25& 1.13 & 2.33 & 0.87 & 2.19 & 1.29 & 3.60 & 1.04 & 3.05 \\[1mm]
& $D$ &$\pi^-$ &  $z, Q^2$ & 25& 1.30 & 1.84 & 1.25 & 1.82 & 1.36 & 4.11 & 1.25 & 3.52 \\[1mm]
\multicolumn{4}{r}{\textbf{SIDIS Total:}}& 110 & 0.97 & 1.64 & 0.78 & 1.55 & 1.30 & 3.97 & 1.11 & 3.40 \\[1mm]
\hline
SLD~\cite{SLD:2003ogn}     & SIA c-tagged &$\pi^{\pm}$ &  $z$ & 21 & 0.77 & 0.61 & 0.84 & 0.61 & 0.73 & 3.00 & 0.81 & 2.55 \\[1mm]
& SIA b-tagged &$\pi^{\pm}$ &  $z$ & 21 & 0.84 & 0.78 & 0.92 & 0.81 & 0.81 & 0.89 & 0.91 & 0.90 \\[1mm]
\multicolumn{4}{r}{\textbf{SIA Total:}} & 42 & 0.80 & 0.69 & 0.88 & 0.71 & 0.77 & 1.95 & 0.86 & 1.73 \\[1mm]
\hline
\multicolumn{4}{r}{\textbf{Total:}} & 152  & 0.92 & 1.38 & 0.81 & 1.31 & 1.15 & 3.41 & 1.04 & 2.94 \\[1mm]
\end{tabular}
\caption{Summary of $\chi^2/N_{pt}$ values for different theoretical setups in the vacuum FFs fits.}
\label{t.ex_vFF}
\end{table*}

The best-fit fragmentation parameters are determined by minimizing the $\chi^{2}$ and then further validated through a series of profile scans on each of those parameters. These parameter space scans are conducted using the \textsc{MINUIT} program~\cite{JAMES1975343}. We apply a tolerance criterion of $\Delta\chi^{2}\sim {\chi^2/N_{pt}}_{max}$ to determine parameter uncertainties. Additionally, we employ the iterative Hessian approach~\cite{Pumplin_2001} to generate error sets of FFs, which can be used to propagate parameter uncertainties to physical observables.


\section{Analysis of Vacuum Fragmentation Functions}
\label{s.vff}

This section presents an extraction of vacuum FFs, which provides the default setup for the subsequent determination of nFFs.
The extraction is performed through a comprehensive study of HERMES data and heavy-flavor-tagged measurements from SLD, as summarized in Table~\ref{t.ex_vFF}.
The HERMES experiment provides SIDIS measurements of hadron production with high precision for charged and neutral $\pi$ over a broad kinematic range, allowing for detailed constraints on vacuum FFs.
Specifically, possible nuclear modifications in the D target data are assumed to be small throughout this analysis, allowing these data to be used as constraints on the vacuum FFs.
Complementary constraints on heavy flavor FFs are provided by SLD data, where heavy-flavor-tagged measurements offer direct sensitivity to $c$ and $b$ quark FFs.

The robustness of our results is systematically evaluated by varying several theoretical setups, and the comparative results are also summarized in Table~\ref{t.ex_vFF}, which provides the $\chi^2/N_{pt}$ values for each data set and for the total fit under different theoretical setups.
We begin by comparing perturbative calculations at NLO and NNLO in QCD to evaluate the impact of higher-order corrections and the stability of the results. The NNLO fits lead to larger $\chi^2$ values than the corresponding NLO fits for the present data set. 
We note that, in the kinematic region covered by the current data, the inclusion of NNLO corrections does not lead to a reduction of scale variations. We therefore use the NLO setup as the default choice in the following analysis. 
The dependence on the initial scale is then examined through comparisons between fits performed with $Q_0 = 1.0$ GeV and $Q_0 = 1.3$ GeV. The latter choice generally leads to an improved description of the data, in particular for the SIDIS measurements. This can be attributed to the positive-definite constraint on the input distributions, because at the lower initial scale DGLAP evolution may favor slightly negative values in certain $z$ regions to improve the fit, which is excluded by construction, whereas the higher $Q_0$ allows a better overall $\chi^2$ within the positive-definite framework.
Furthermore, the role of theoretical uncertainties introduced through the covariance matrix (see Section~\ref{s.goodness}) is investigated by comparing fits with and without their inclusion. While including theoretical uncertainties leads to reduced $\chi^2$ values, the reduction does not correspond to a significant improvement in the fit quality. Given the current experimental precision and the already satisfactory fit quality, the inclusion of such uncertainties is therefore not required in the present analysis.
Based on these comparisons, we identify a default configuration characterized by $Q_0 = 1.3$ GeV, NLO perturbative calculations, and the exclusion of theoretical uncertainties in the fitting procedure. This setup provides a stable and balanced description of the data across different data sets.
For this default choice, the total $\chi^2$ amounts to 158.2 for a total of 152 data points, corresponding to $\chi^2/N_{pt} = 1.04$, indicating good overall agreement between theory and experiment. The corresponding $\chi^2/N_{pt}$ values are 1.11 and 0.86 for the SIDIS and SIA data sets, respectively.

We now turn to the details of the vacuum FFs extraction. Section~\ref{ss.ep} focuses on comparisons between theoretical predictions and experimental data for SIDIS and SIA observables, while Section~\ref{ss.vff} presents the extracted vacuum FFs and investigates their dependence on different theoretical configurations.

\subsection{Comparison with Experimental Data}
\label{ss.ep}

\begin{figure*}[t]
\centering
\includegraphics[width = 0.9 \textwidth]{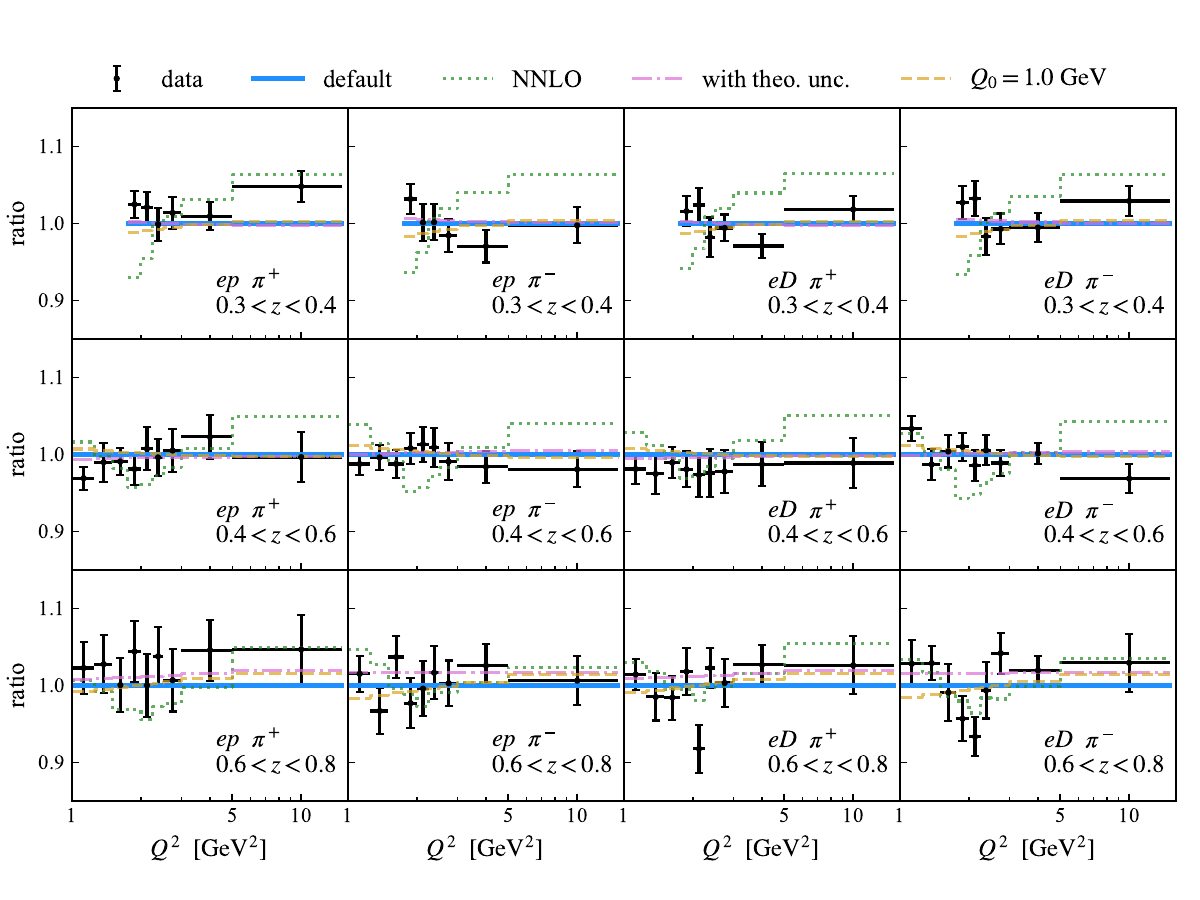}
\caption{Comparison of HERMES measurements~\cite{HERMES:2012uyd} and theoretical predictions for $\pi^+$ and $\pi^-$ production in SIDIS off $p$ and D targets, shown in bins of $z$ as functions of $Q^2$. All results are presented as ratios to the default prediction, defined as the setup with $Q_0 = 1.3$ GeV, NLO, and the exclusion of theoretical uncertainties. The black points with error bars represent the experimental data divided by the default prediction. The solid blue line corresponds to the default prediction and is therefore equal to unity by construction. The dotted green line shows the NNLO prediction, the dash-dotted pink line shows the result obtained when theoretical uncertainties are included in the fit, and the dashed orange line corresponds to the initial scale choice $Q_0=1.0~\mathrm{GeV}$.} 

\label{f.vff_pip_pim}
\end{figure*}

\begin{figure*}[t]
\centering
\includegraphics[width = 0.49 \textwidth]{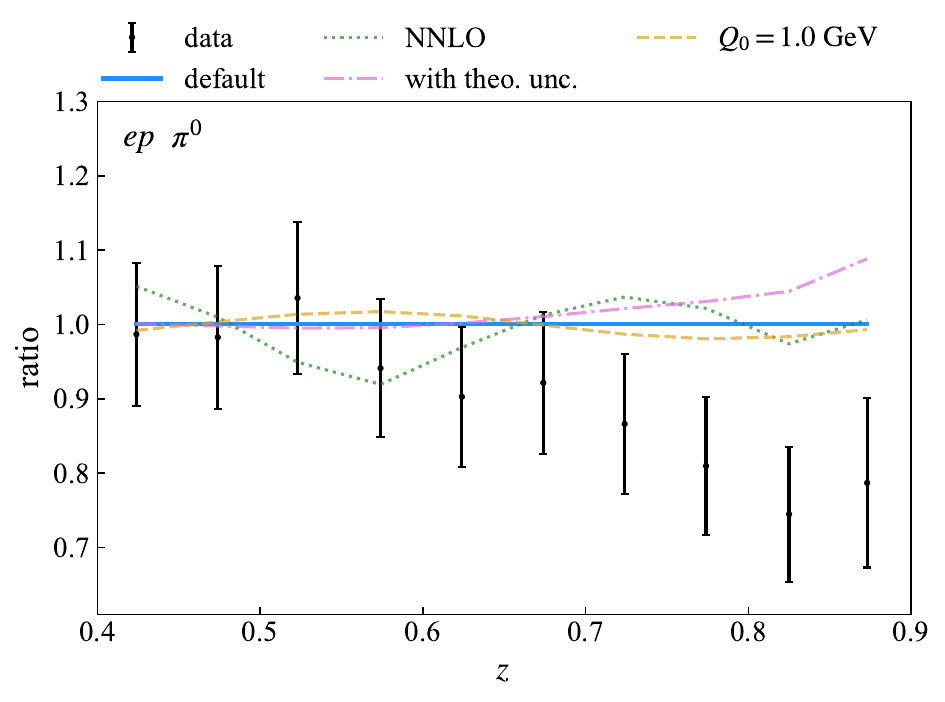}
\caption{
Similar to Fig.~\ref{f.vff_pip_pim}, but for $\pi^0$ production in HERMES experiments~\cite{HERMES:2001ghm}, shown as a function of $z$.}
\label{f.vff_hermes_pi0}
\end{figure*}

\begin{figure*}[t]
\centering
\includegraphics[width = 0.9 \textwidth]{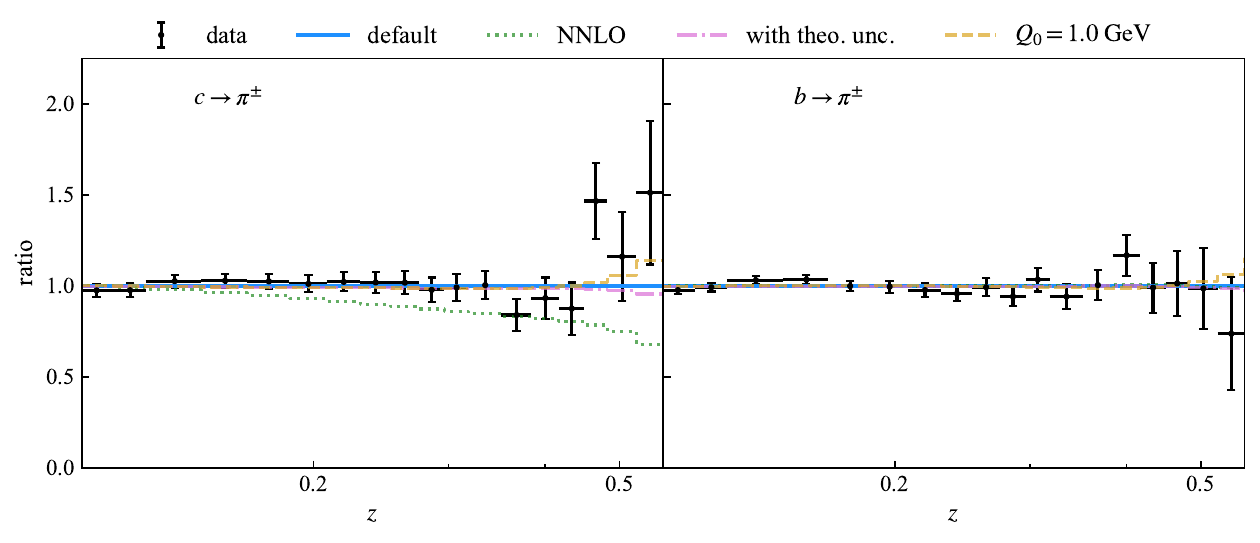}
\caption{Similar to Fig.~\ref{f.vff_pip_pim}, but for $\pi^{\pm}$ production in
SIA at SLD~\cite{SLD:2003ogn}, shown as a function of $z$. }
\label{f.vff_pisum}
\end{figure*}

Figure~\ref{f.vff_pip_pim} compares the HERMES measurements~\cite{HERMES:2012uyd} with theoretical predictions for $\pi^\pm$ production in SIDIS off $p$ and D targets under several theoretical setups considered in this work. To facilitate a direct comparison among the different theoretical predictions and the experimental data, both the experimental data and the theoretical predictions are divided by the default prediction in each kinematic bin.
The experimental uncertainties are small, remaining at the level of $2\%$-$4\%$ for $z<0.6$ and increasing moderately to below $5\%$ in the range $0.6<z<0.8$. This high precision provides stringent constraints on the extracted vacuum FFs.
The default theoretical prediction describes the $p$ data well across the full kinematic range. A mild deviation is observed in the $\pi^+$ channel for $0.3<z<0.4$, where the data lie approximately $5\%$ above the central prediction. Although this deviation exceeds the experimental uncertainties, its impact on the fit quality remains limited because only a small number of data points are affected.
For the D target, the data show some mild fluctuations relative to the
default prediction. Since these deviations remain compatible with the
experimental uncertainties, the default prediction still provides a
consistent description of the data.
The combined fit quality for $p$ and D datasets demonstrates the stability of the extraction and supports the reliability of the resulting vacuum FFs.
The impact of the theoretical setup is then assessed by comparing the different predictions. In the low and intermediate $z$ region ($z<0.6$), the default setup, the fit including theoretical uncertainties, and the initial scale choice $Q_0=1.0~\mathrm{GeV}$ give nearly identical predictions. This indicates that the extracted vacuum FFs are stable with respect to these variations. A different pattern becomes visible at larger $z$ ($0.6<z<0.8$), where the $Q_0=1.0$ GeV setup lies slightly below the default at low $Q^2$ and slightly above it at higher $Q^2$, while the setup including theoretical uncertainties shows a mild upward shift over the full $Q^2$ range. The NNLO calculation shows qualitatively different behavior across all $z$ regions. In the region $0.3<z<0.4$, the NNLO prediction evolves from a suppression at low $Q^2$ (approximately $5\%$ below the default) to an enhancement at high $Q^2$.

\begin{figure*}[t]
\centering
\includegraphics[width = 1.0 \textwidth]{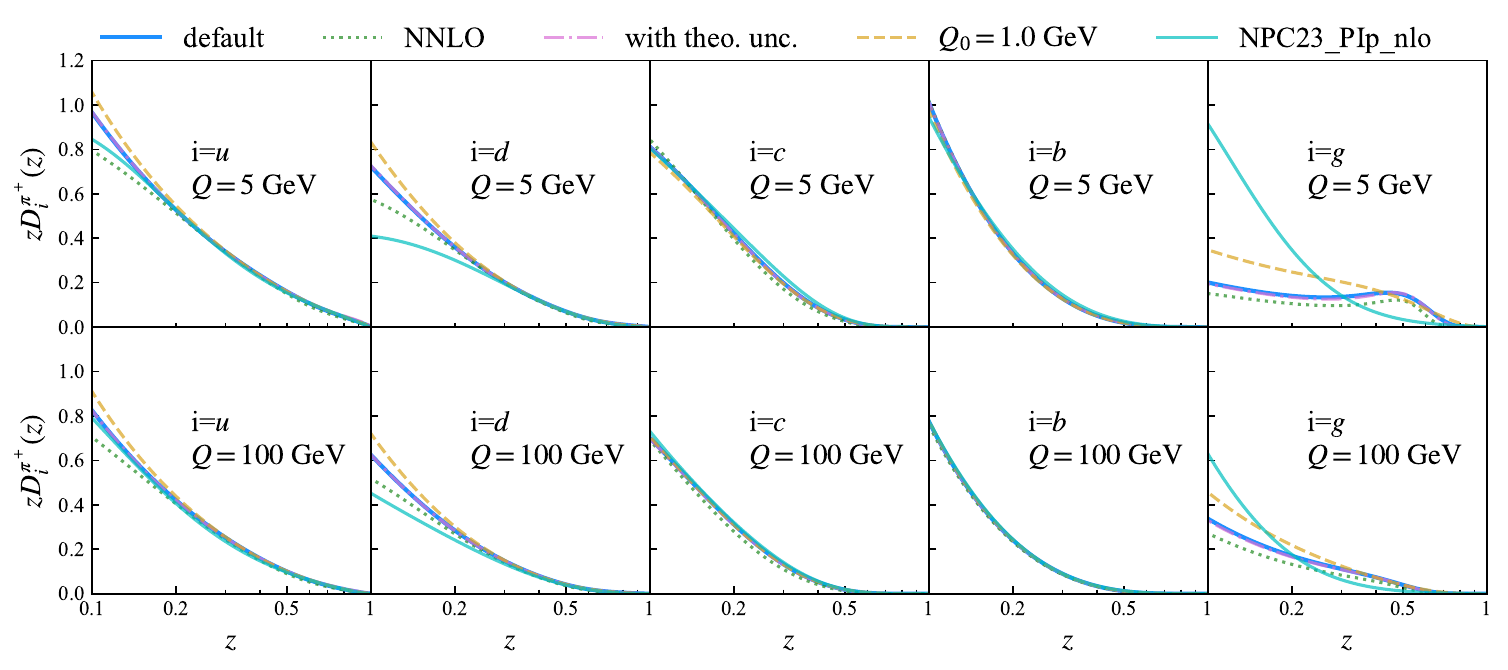}
\caption{
Comparison of the extracted vacuum FFs for $\pi^+$ with the NPC23~\cite{Gao_2024} for different parton flavors at $Q=5$ GeV and $Q=100$ GeV.}
\label{f.vff}
\end{figure*}

For $\pi^{0}$ production, as shown in Fig.~\ref{f.vff_hermes_pi0}, the agreement between theory and data remains good, with a mild overestimation observed in the high-$z$ region ($z>0.7$).
This feature persists across all theoretical configurations considered, indicating that it is not driven by perturbative order or initial scale choices. 
Such deviations may reflect limitations of the current theoretical framework in describing $\pi^{0}$ production at high $z$, potentially requiring modifications of FFs near $z \to 1$.
Nevertheless, given the relatively large uncertainties and the limited statistical precision of the $\pi^{0}$ data, this deviation does not significantly impact the total fit and remains consistent with the overall description of the data.

Figure~\ref{f.vff_pisum} presents the comparison between theory and data for charm tagged and bottom tagged $\pi^\pm$ production in SIA at the $Z$ pole, based on measurements from the SLD Collaboration. These data provide clean probes of heavy quark fragmentation due to their high flavor purity and well-controlled kinematics. For clarity of presentation, the error bars of the last three data points in charm tagged and the last two data points in bottom tagged are not displayed in the figure, as the experiment uncertainties become large after normalization.
The default theoretical prediction describes the data well across most of the $z$ range. Although mild deviations are observed at large $z$ ($z \gtrsim 0.5$), where the last data points slightly exceed the experimental uncertainty, this behavior remains compatible with the data within the corresponding large experimental uncertainties. The predictions obtained using the NNLO setup and those obtained with $Q_0 = 1.0~\mathrm{GeV}$ are close to the default results over the full $z$ range, except for the charm-tagged data, where the NNLO predictions are found to lie below the default results.

\subsection{Extracted Vacuum Fragmentation Functions}
\label{ss.vff}

The extracted default vacuum FFs for gluon and quarks are shown in Fig.~\ref{f.vff} at scales $Q=5$ GeV and $Q=100$ GeV. 
To assess the impact of different theoretical assumptions, we compare results obtained under several configurations, including NNLO corrections, a lower initial scale ($Q_0=1.0$ GeV), and the inclusion of theoretical uncertainties. 
In addition, a comparison with the NPC23 FFs~\cite{Gao_2024}, which are constrained by high-energy data, is included as a reference to assess the consistency of our results.
It should be noted that the experimental data included in our analysis are primarily restricted to the region $z > 0.3$.

For the light quarks $u$ and $d$, all configurations show good agreement for $z>0.2$. 
Differences appear mainly at low $z$, where the NNLO result is slightly suppressed, while the $Q_0=1.0$ GeV configuration shows a mild enhancement. 
The inclusion of theoretical uncertainties has a negligible impact on the central values. 
Compared to our extraction, the NPC23 distributions are generally lower in the low $z$ region.
The heavy quark $c$ and $b$ FFs exhibit a high degree of consistency across all configurations and with NPC23. 
This robustness reflects the strong constraints provided by the SLD heavy-flavor-tagged measurements at the $Z$ pole, which effectively fix the normalization and shape of heavy quark fragmentation.
For the gluon distribution, the configurations considered in this work exhibit very similar overall shapes. The low $z$ region ($z<0.3$) shows a decrease, whereas a pronounced local maximum is observed around $z\approx 0.5$.
Despite this similarity in shape, noticeable differences in magnitude arise across different configurations. The $Q_0=1.0$ GeV setup leads to a significant enhancement in the low $z$ region, while the NNLO calculation produces a systematically suppressed distribution over the full $z$ range.
The NPC23 result exhibits a decrease over the full $z$ range and is significantly larger than our result at low $z$, while becoming smaller around $z=0.3$. 

\begin{table*}[t]
\centering
\small
~\\
\setlength{\tabcolsep}{4pt}
\begin{tabular}{cccccccccccccc}
 Exp & Target(A) & Particle & binning & $N_{pt} $&\multicolumn{3}{c}{nCTEQ($\chi^2/N_{pt}$)} &  \multicolumn{3}{c}{nNNPDF($\chi^2/N_{pt}$)} & \multicolumn{3}{c}{EPPS21($\chi^2/N_{pt}$)} \\
&& & & &$\delta=0$&$\delta=0.5$&$\delta=1$&$\delta=0$&$\delta=0.5$&$\delta=1$&$\delta=0$&$\delta=0.5$&$\delta=1$\\
\hline\hline
HERMES~\cite{HERMES:2001ghm}   & $p$ &$\pi^0$ &  $z$ & 10 & 2.30 & 2.57 & 2.53 & 2.22 & 2.55 & 2.50 & 2.23 & 2.56 & 2.52 \\[1mm]
HERMES~\cite{HERMES:2007plz}  & $He$ &$\pi^{\pm}$ &  $z, \nu$ & 16 & 0.23 & 0.25 & 0.25 & 0.21 & 0.24 & 0.24 & 0.23 & 0.26 & 0.26 \\[1mm]
& $He$ &$\pi^{\pm}$ &  $z, Q^2$& 20 & 0.20 & 0.25 & 0.25 & 0.20 & 0.25 & 0.25 & 0.21 & 0.25 & 0.25 \\[1mm]
& $He,Ne,Kr,Xe$ &$\pi^+$ &  $z$& 24 & 0.31 & 0.61 & 1.44 & 0.42 & 0.60 & 1.35 & 0.33 & 0.59 & 1.44 \\[1mm]
& $He,Ne,Kr,Xe$ &$\pi^-$ &  $z$& 24 & 0.17 & 0.44 & 0.99 & 0.50 & 0.75 & 1.40 & 0.18 & 0.44 & 1.05 \\[1mm]
& $He,Ne,Kr,Xe$ &$\pi^0$ &  $z$& 24 & 0.28 & 0.54 & 1.14 & 0.28 & 0.47 & 1.10 & 0.28 & 0.51 & 1.16 \\[1mm]
HERMES~\cite{HERMES:2011qjb}   & $Ne,Kr,Xe$ &$\pi^+$ &  $z, Q^2$ & 60 & 0.49 & 2.37 & 2.44 & 0.65 & 1.83 & 1.76 & 0.44 & 1.96 & 2.01 \\[1mm]
& $Ne,Kr,Xe$ &$\pi^-$ &  $z, Q^2$ & 60 & 0.70 & 2.29 & 2.41 & 0.76 & 2.22 & 2.38 & 0.63 & 2.03 & 2.11 \\[1mm]
& $Ne,Kr,Xe$ &$\pi^+$ &  $z, \nu$ & 51 & 2.47 & 1.11 & 2.41 & 1.88 & 1.02 & 2.84 & 2.13 & 1.03 & 2.60 \\[1mm]
& $Ne,Kr,Xe$ &$\pi^-$ &  $z, \nu$ & 51 & 2.27 & 0.96 & 1.40 & 2.52 & 1.09 & 1.60 & 2.06 & 0.86 & 1.47 \\[1mm]
HERMES~\cite{HERMES:2012uyd}   & $p$ &$\pi^+$ &  $z, Q^2$ & 25 & 1.17 & 1.20 & 1.32 & 1.08 & 1.03 & 1.11 & 1.13 & 1.09 & 1.19 \\[1mm]
& $p$ &$\pi^-$ &  $z, Q^2$ & 25 & 0.83 & 0.77 & 0.90 & 0.81 & 0.96 & 1.14 & 0.84 & 0.88 & 1.04 \\[1mm]
& $D$ &$\pi^+$ &  $z, Q^2$ & 25 & 1.25 & 1.21 & 1.28 & 1.36 & 1.16 & 1.16 & 1.33 & 1.20 & 1.24 \\[1mm]
& $D$ &$\pi^-$ &  $z, Q^2$ & 25 & 1.33 & 1.16 & 1.19 & 1.17 & 1.19 & 1.28 & 1.32 & 1.19 & 1.24 \\[1mm]
CLAS~\cite{CLAS:2021jhm}   & $C,Fe,Pb$ &$\pi^+$ &  $z, Q^2, \nu$ & 117 & 0.83 & 0.85 & 1.47 & 0.81 & 0.84 & 1.56 & 0.82 & 0.83 & 1.45 \\[1mm]
& $C,Fe,Pb$ &$\pi^-$ &  $z, Q^2, \nu$ & 114 & 1.99 & 1.62 & 1.60 & 2.04 & 1.65 & 1.58 & 2.02 & 1.66 & 1.64 \\[1mm]
\multicolumn{4}{r}{\textbf{SIDIS Total:}} & 671 & 1.19 & 1.27 & 1.60 & 1.20 & 1.23 & 1.61 & 1.15 & 1.20 & 1.56\\[1mm]
\hline
SLD~\cite{SLD:2003ogn}     & c-tagged &$\pi^{\pm}$ &  $z$ & 21 & 0.68 & 0.87 & 0.90 & 0.72 & 0.88 & 0.92 & 0.67 & 0.85 & 0.89 \\[1mm]
& b-tagged &$\pi^{\pm}$ &  $z$ & 21 & 0.81 & 0.89 & 0.90 & 0.83 & 0.90 & 0.90 & 0.81 & 0.89 & 0.90 \\[1mm]
\multicolumn{4}{r}{\textbf{SIA Total:}} & 42 & 0.75 & 0.88 & 0.90 & 0.78 & 0.89 & 0.91 & 0.74 & 0.87 & 0.90\\[1mm]
\hline
\multicolumn{4}{r}{\textbf{Total:}} & 713 & 1.17 & 1.24 & 1.56 & 1.17 & 1.21 & 1.57 & 1.12 & 1.18 & 1.52 \\[1mm]
\end{tabular}
\caption{Per-dataset and total $\chi^2/N_{pt}$ values for all datasets using different nuclear PDF parameterizations (nCTEQ15WZ, nNNPDF, EPPS21) with various $\delta=0$, $\delta=0.5$, $\delta=1$ schemes.}
\label{t.chi2_nFF}
\end{table*}


\section{Analysis of Nuclear Fragmentation Functions}
\label{s.nff}

Based on the investigation of theoretical setups, we proceed to the extraction of nFFs. Compared to the vacuum FFs, the determination of nFFs is primarily based on SIDIS data from the HERMES and CLAS experiments.
The datasets include a variety of nuclear targets, such as He, Ne, Kr, Xe, C, and Pb, spanning a broad range of atomic mass numbers, while the measurements are presented in various kinematic bins, including $z$, $(z,\nu)$, $(z,Q^2)$, and $(z,\nu,Q^2)$, providing multidimensional constraints on the hadronization process.

We adopt the default theoretical setup established in the previous section at NLO perturbative accuracy, without theoretical uncertainties. The initial scale is set to $Q_0 = 1.3\,\mathrm{GeV}$.
Within this framework, vacuum FFs and nuclear modifications are fitted simultaneously, with the $\nu$ dependence incorporated as described in Section~\ref{s.theory}.
To study the impact of the $\nu$ dependence, we consider three representative choices of $\delta = 0, 0.5, 1$ in Eq.~\eqref{eq.fa}. The sensitivity of the extracted nFFs to the assumed $\nu$ dependence is quantified by comparing the fit quality and the resulting variations across these scenarios.

To further assess the uncertainties arising from initial-state nuclear effects, we supplement the default nPDF set nCTEQ15WZ~\cite{Kusina:2020lyz} with two recent global nPDF analyses: nNNPDF3.0~\cite{AbdulKhalek:2022fyi} and EPPS21~\cite{Eskola:2021nhw}. By repeating the nFFs extraction procedure with these three nPDF sets, we examine the stability of the extracted nFFs with respect to the choice of nPDFs.

\subsection{Comparison with Nuclear Data}
\label{ss.nff_ep}

Table~\ref{t.chi2_nFF} summarizes the quality of the fits, quantified by $\chi^2/N_{pt}$, for different choices of nPDF sets and $\nu$ dependence. 
Since the dependence on the nPDF input is found to be weak, we first discuss this point to justify the use of the nCTEQ15WZ results in the following figures. The $\chi^2/N_{pt}$ values for each individual dataset show a similar level of agreement across nCTEQ15WZ, nNNPDF, and EPPS21, indicating that the extraction of nFFs is not strongly affected by the specific implementation of initial state nuclear effects. To further illustrate this feature, a dedicated benchmark study is presented in Appendix~\ref{appendix}, where the default nFFs extracted using the nCTEQ15WZ set are kept fixed while the SIDIS cross sections are recalculated with nNNPDF3.0 and EPPS21. This weak dependence may be attributed to the use of hadron multiplicities, defined in Eq.~\eqref{eq.multi} as the ratio of the semi-inclusive hadron-production cross section to the inclusive DIS cross section. Initial-state nuclear effects from nPDFs enter both the numerator and denominator of this ratio and tend to cancel. For clarity in the following comparison, only the results obtained with the nCTEQ15WZ are shown in the figures. 
Having established that the nPDF-set dependence is small, we now focus on the sensitivity to the assumed $\nu$ dependence. The results show that the $\delta=1$ scheme generally leads to a significantly worse description of the data across all nPDF sets, as indicated by the larger $\chi^2/N_{pt}$ values. In contrast, the $\delta=0$ and $\delta=0.5$ schemes both improve the description relative to $\delta=1$ and give comparable fit qualities. The total $\chi^2$ values alone therefore do not establish a clear preference between these two scenarios. A more differential comparison is provided by the HERMES~\cite{HERMES:2007plz} measurements, where the $\nu$ dependence of $R_A^h$ can be examined in fixed $z$ regions, as discussed below.

\begin{figure*}[t]
\centering
\begin{overpic}[width = 0.65 \textwidth]{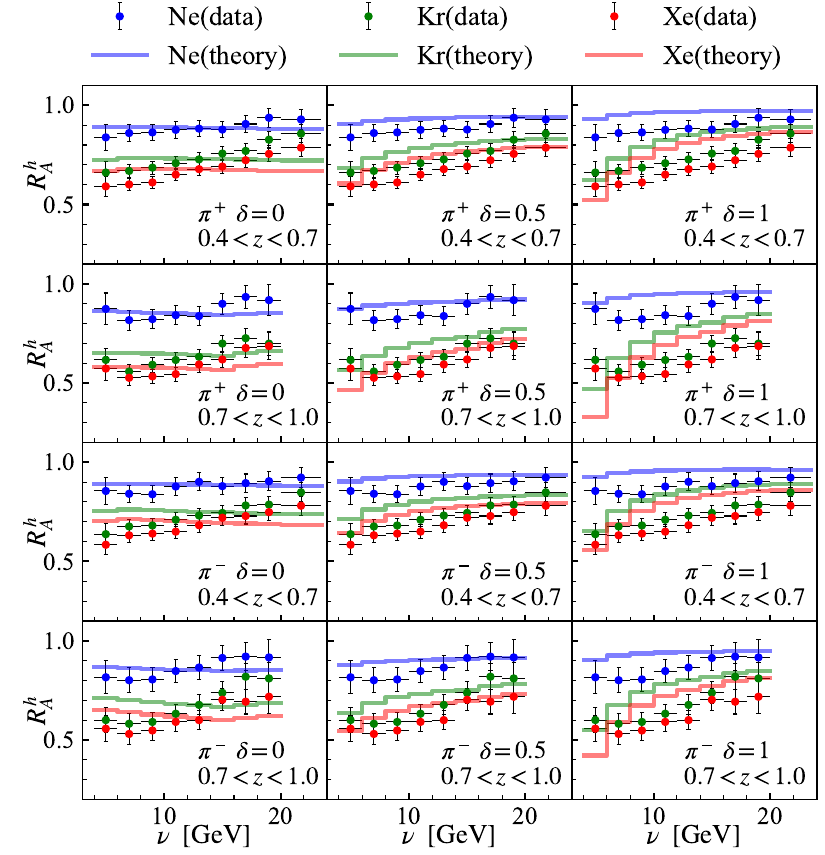}
\end{overpic}
\caption{
Comparison of the measured and predicted nuclear modification with respect to $z$ and $\nu$ for $\pi^+$ and $\pi^-$ in HERMES experiments~\cite{HERMES:2011qjb}.
}
\label{f.nff_iunc0_CTEQ_hermes_pi_4}
\end{figure*}

\begin{figure*}[t]
\centering

\begin{overpic}[width=0.7\textwidth]{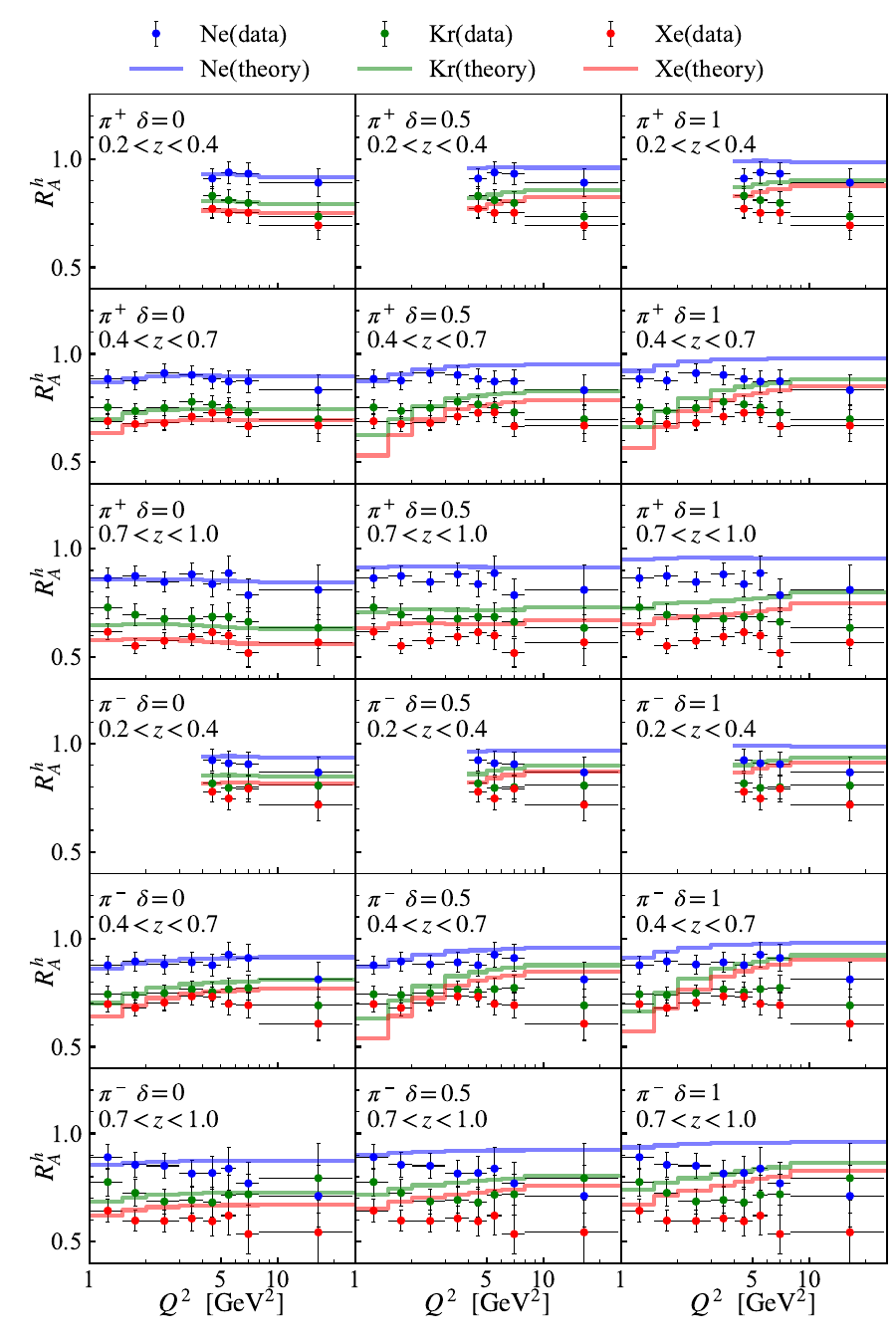}
\end{overpic}

\caption{
Similar to Fig.~\ref{f.nff_iunc0_CTEQ_hermes_pi_4}, but shown with
respect to $z$ and $Q^2$.
}
\label{f.nff_iunc0_CTEQ_hermes_pi_23}
\end{figure*}

\begin{figure*}[ht]
\centering
\begin{overpic}[width = 0.9 \textwidth]{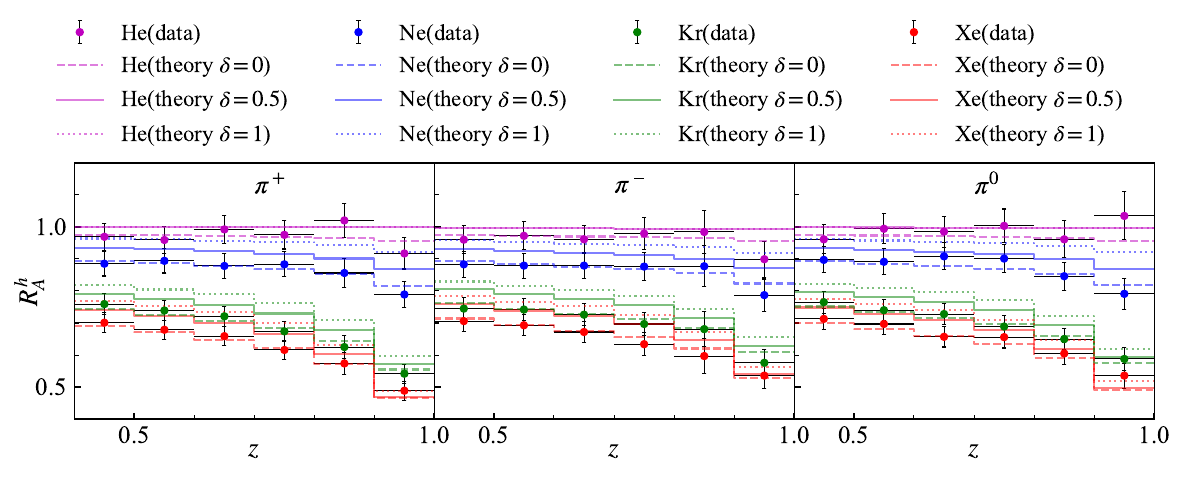}
\end{overpic}
\caption{
Comparison of the measured and predicted nuclear modification with respect to $z$ for $\pi$ production in HERMES experiments~\cite{HERMES:2007plz}. Theoretical predictions are obtained for different $\nu$ dependences and curves.
}
\label{f.nff_iunc0_CTEQ_hermes_pi_1}
\end{figure*}

\begin{figure*}[t]
\centering

\begin{overpic}[width=0.7\textwidth]{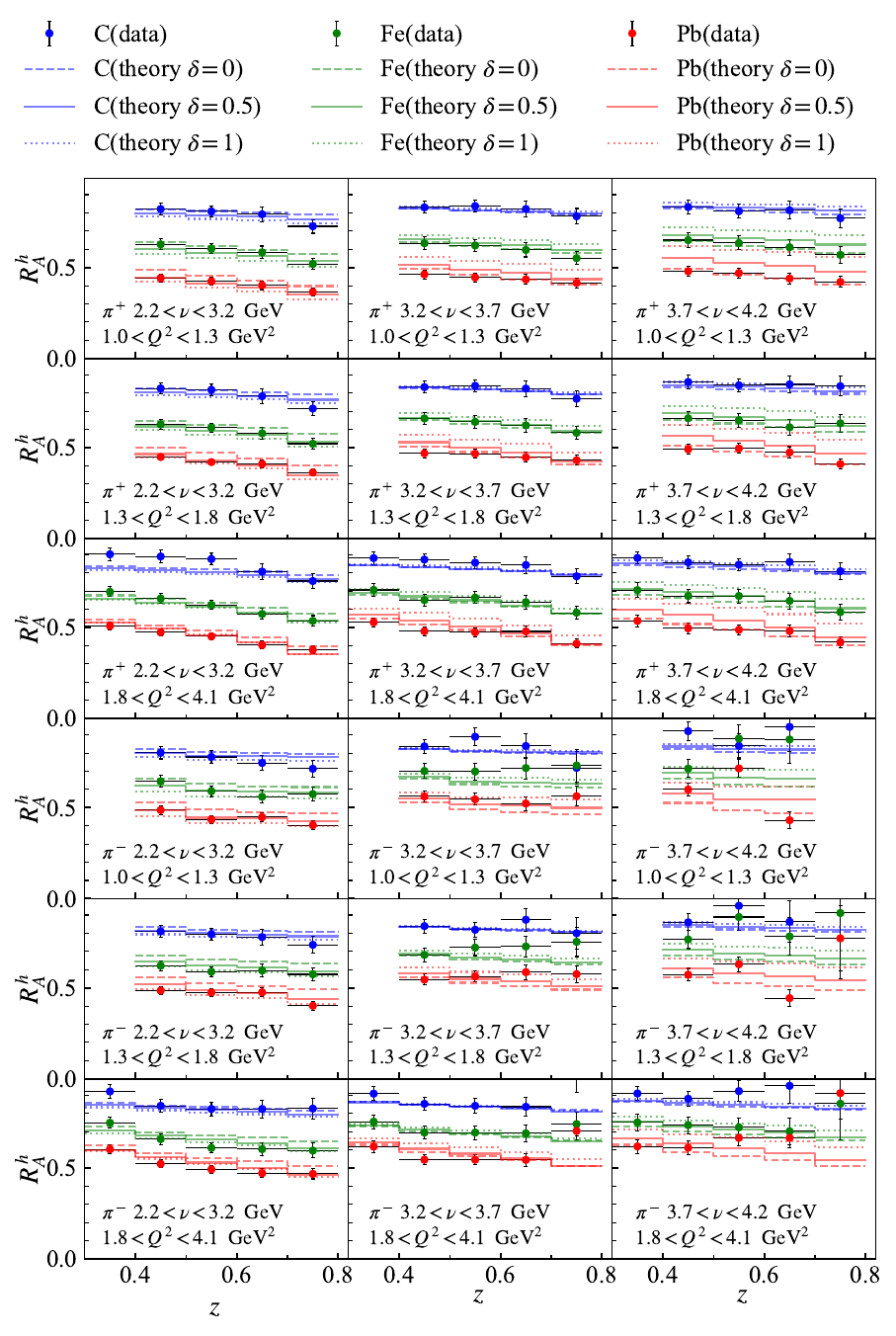}
\end{overpic}

\caption{
Comparison of the measured and predicted nuclear modification with
respect to $z$ and $\nu$ for $\pi^+$ and $\pi^-$ in CLAS experiments~\cite{CLAS:2021jhm}.
Theoretical predictions are shown for different assumptions on the
$\nu$ dependence. The C, Fe, and Pb targets are shown in blue, green, and
red, respectively. The $\delta=0.5$, $\delta=0$, and $\delta=1$ schemes
are represented by solid, dashed, and dotted lines, respectively.
}
\label{f.nff_iunc0_CTEQ_clas_pi}
\end{figure*}

Figure~\ref{f.nff_iunc0_CTEQ_hermes_pi_4} presents the HERMES~\cite{HERMES:2007plz} measurements of the nuclear modification ratio $R_A^h$ as a function
of $\nu$ for $\pi^+$ and $\pi^-$ production on Ne, Kr, and Xe targets. The results are shown in two representative $z$ regions, and the three $\nu$-dependent parametrizations are displayed separately. This representation provides a direct test of the assumed $\nu$ dependence in the nFFs.
A clear increase of $R_A^h$ with $\nu$ is observed, especially for the heavier Kr and Xe and in the larger $z$ region. This behavior indicates that the nuclear modification becomes weaker as the energy transferred to the fragmenting parton increases. The $\delta=0$ scheme, which contains no explicit $\nu$ dependence in the nuclear modification, therefore does not capture the observed rise as well and tends to give a flatter behavior. In contrast, the $\delta=1$ scheme generates a stronger increase with $\nu$, but it generally overestimates the data, in particular for heavier nuclei and at larger $z$. The $\delta=0.5$ scheme provides the most balanced description. It captures the increasing trend with $\nu$ while keeping the overall magnitude of $R_A^h$ close to the data.


Figure~\ref{f.nff_iunc0_CTEQ_hermes_pi_23} shows the HERMES~\cite{HERMES:2007plz} measurements as a function of $Q^2$ for $\pi^+$ and $\pi^-$ production on Ne, Kr, and Xe targets in different $z$ bins. 
The data show the expected nuclear mass dependence, with the suppression becoming stronger from Ne to Kr and Xe. The overall dependence on
$Q^2$ is relatively mild in all three $z$ intervals, while the suppression becomes more pronounced as $z$ increases. For $\pi^+$ production, both the $\delta=0$ and $\delta=0.5$ schemes provide a reasonable description of the data, considering the relatively
large experimental uncertainties, especially in the large-$z$ region and
at high $Q^2$. The $\delta=0.5$ predictions are slightly higher in some bins, particularly for heavier nuclei and at larger $z$, while the $\delta=1$ scheme tends to overestimate the data. For $\pi^-$ production, similar features are observed. The data show a weak $Q^2$ dependence, while the suppression becomes stronger from Ne to Kr and Xe. Compared with the $\pi^+$ case, the $\pi^-$ data exhibit more visible variations, especially in the intermediate and large $z$ intervals. The $\delta=0$ and $\delta=0.5$ schemes both remain compatible with the measurements, while the $\delta=1$ scheme tends to give higher $R_A^h$ values, especially for the heavier targets.

Figure~\ref{f.nff_iunc0_CTEQ_hermes_pi_1} presents the 2007 HERMES~\cite{HERMES:2007plz} measurements of the nuclear modification ratio $R_A^h$ for $\pi^+$, $\pi^-$, and $\pi^0$ on He, Ne, Kr, and Xe targets. 
For each nuclear target, the $z$ distributions of $\pi^+$, $\pi^-$, and $\pi^0$ are largely similar in shape.
The He data remain close to unity over the measured $z$ range, showing that nuclear effects are weak for the lightest target. For Ne, Kr, and Xe, a clear suppression pattern develops with increasing $z$, becoming stronger for heavier nuclei, which provides important experimental input for the determination of the $A$ dependence in the present analysis.
From the figure, we observe that all three $\nu$ schemes provide a satisfactory description of the data, supporting a stable and consistent extraction.

Figure~\ref{f.nff_iunc0_CTEQ_clas_pi} shows the nuclear modification ratio $R_A^h$ for $\pi^+$ and $\pi^-$ measured by the CLAS collaboration. 
A clear overall trend is observed, with $R_A^h$ decreasing as a function of $z$. 
For $\pi^+$ production, the $Q^2$ dependence can be seen by comparing the $z$ distributions across different $Q^2$ bins within the same $\nu$ interval, with higher $Q^2$ values leading to an upward shift of the $R_A^h$ distributions. This effect becomes less pronounced for heavier nuclei, suggesting a reduced sensitivity to $Q^2$ as the mass number $A$ increases.
The $\nu$ dependence is inferred by comparing the $z$ distributions across different $\nu$ bins within the same $Q^2$ interval. In the low-$z$ region, the data exhibit only a weak dependence on $\nu$, while at larger $z$ an enhancement with increasing $\nu$ becomes visible. 
A comparison among different $\nu$ schemes shows that the $\delta=0$ scheme, which gives the same nuclear modification at fixed $z$ and $Q^2$ when only $\nu$ is varied, slightly deviates from the data in the lowest $\nu$ interval but performs reasonably well elsewhere. The predictions of the $\delta=0.5$ scheme provide a good overall description of the data, with minor overestimation observed at higher $\nu$. Meanwhile, the $\delta=1$ scheme tends to overestimate the data at intermediate and large $\nu$, consistent with its larger $\chi^2$ value reported in Table~\ref{t.chi2_nFF}. 
For $\pi^-$ production, however, more pronounced structures emerge at larger $z$ ($z>0.5$). In particular, peak-like features develop in the $3.2<\nu<3.7$ GeV range, becoming more visible at $3.7<\nu<4.2$ GeV range. These structures are most prominent in the lower $Q^2$ bins and are accompanied by increasing experimental uncertainties.
Despite these localized features, the predictions of the $\delta=0.5$ scheme provide a satisfactory description of the $\pi^-$ data across $2.2<\nu<3.2$ GeV and $3.2<\nu<3.7$ GeV regions. In the higher-$\nu$ range ($3.7<\nu<4.2$ GeV), all three $\nu$ parametrizations show a relatively poor description of the data. However, given the relatively large experimental uncertainties in this region, the $\chi^2/N_{pt}$ values of CLAS $\pi^-$ remain within a reasonable range.

\begin{figure*}[t]
\centering
\includegraphics[width = 0.9 \textwidth]{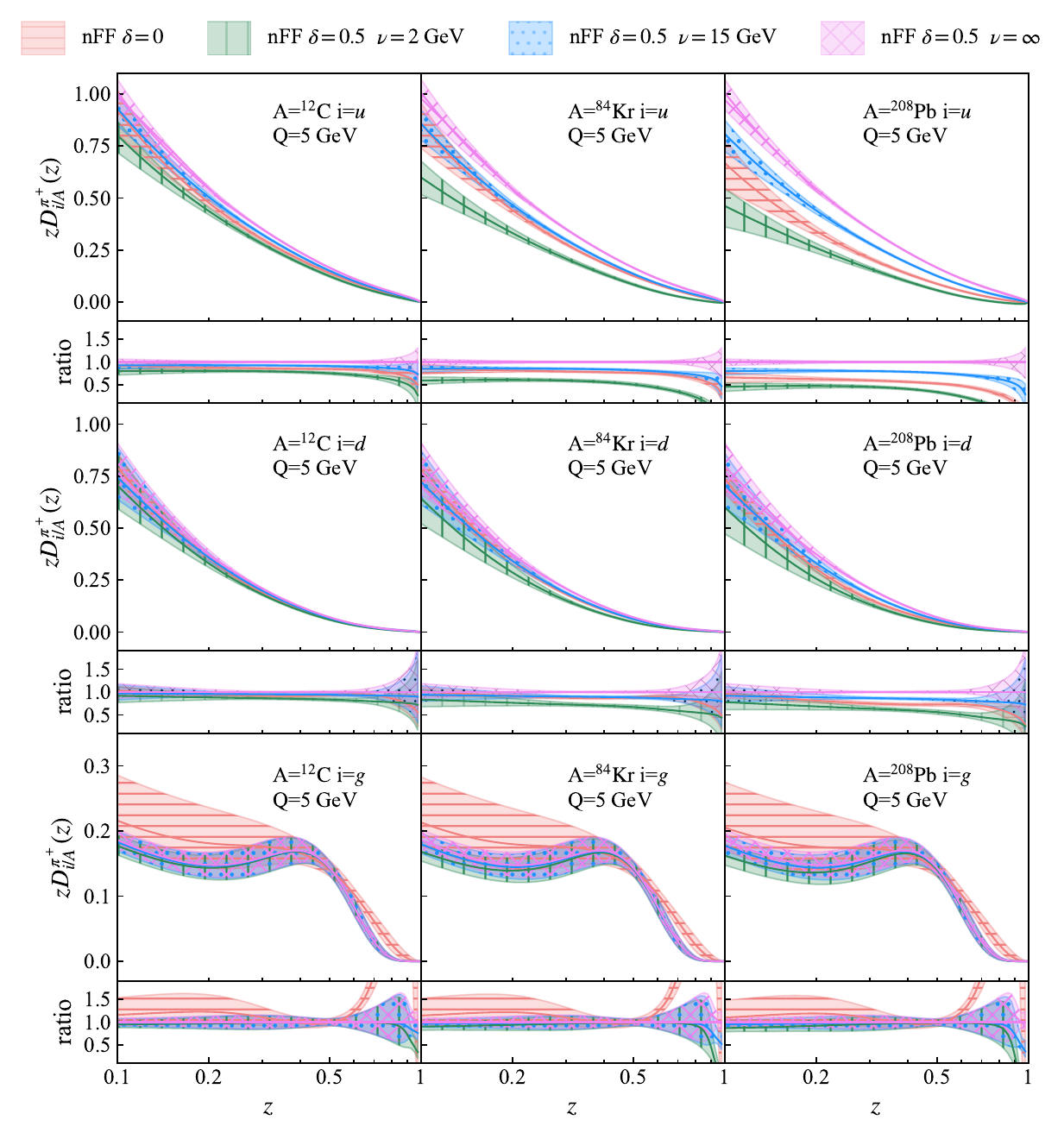}
\caption{
Extracted nFFs for $\pi^+$ obtained with the nCTEQ15WZ nPDFs for $u$ quarks, $d$ quarks, and gluons on C, Kr, and Pb targets at $Q=5$ GeV, and ratios normalized to the case with $\delta=0.5$ and $\nu=\infty$. The results are shown for the default $\delta=0.5$ scheme at $\nu=2~\mathrm{GeV}$, $\nu=15~\mathrm{GeV}$, and $\nu\to\infty$, together with the $\nu$-independent result from the $\delta=0$ scheme. The Hessian uncertainty bands are also included.
}
\label{hessian_nCTEQ_5_GeV}
\end{figure*}

\subsection{Extracted Nuclear Fragmentation Functions}
\label{ss.nff_modification}

In this subsection, we first present the extracted nFFs obtained within the $\delta=0.5$ scheme, which incorporates the dependence of both $A$ and $\nu$. 
Figures~\ref{hessian_nCTEQ_5_GeV} summarize the results at the scales $Q = 5$ GeV for $u$ quarks, $d$ quarks, and gluons on C, Kr, and Pb targets, which allows for a direct comparison of different flavors at fixed $A$, as well as nFFs with increasing $A$ for a given flavor. For the default $\delta=0.5$ scheme, each panel shows the results at three representative values of the energy transfer, $\nu=2~\mathrm{GeV}$, $\nu=15~\mathrm{GeV}$, and $\nu\to\infty$. In this scheme, the $\nu\to\infty$ limit corresponds to the vacuum FFs because $\mathcal{F}_A(A,\nu)$ vanishes for $\delta>0$. For comparison, the $\nu$-independent result obtained with the $\delta=0$ scheme is also shown.

In the region $z > 0.1$, the nuclear modifications for all parton species are reasonably well constrained by the available experimental data. 
For $u$ and $d$ quarks, the curves for $\nu = 2$~GeV, $\nu = 15$~GeV, and $\nu \to \infty$ show a clear separation, with the prediction of vacuum FFs part lying above the $\nu = 15$~GeV result, which in turn remains above the $\nu = 2$~GeV curve.
This behavior follows naturally from the $\nu$-dependent structure of the parametrization in the $\delta = 0.5$ scheme, where smaller values of $\nu$ lead to stronger suppression for the same $A$.
The difference becomes more pronounced with increasing mass number $A$. 
The nuclear suppression of the $u$ quark is found to reach about $20\%$ for a C nucleus at $\nu=2~\mathrm{GeV}$, and increases to nearly $50\%$ for Pb.
A similar trend is observed for the $d$ quark, where the suppression is about $20\%$ for C at $\nu=2~\mathrm{GeV}$ and rises to nearly $30\%$ for Pb. 
This indicates that the sensitivity of nuclear medium effects to the variable $\nu$ is enhanced in heavier nuclei.
For gluons, the curves for $\nu=2~\mathrm{GeV}$, $\nu=15~\mathrm{GeV}$,
and $\nu\to\infty$ are very close to each other. This behavior arises
because the gluon nuclear modifications are set to zero at the initial
scale and are generated only through QCD evolution.

\begin{figure*}[!t]
\centering
\begin{overpic}[width = 0.49 \textwidth]{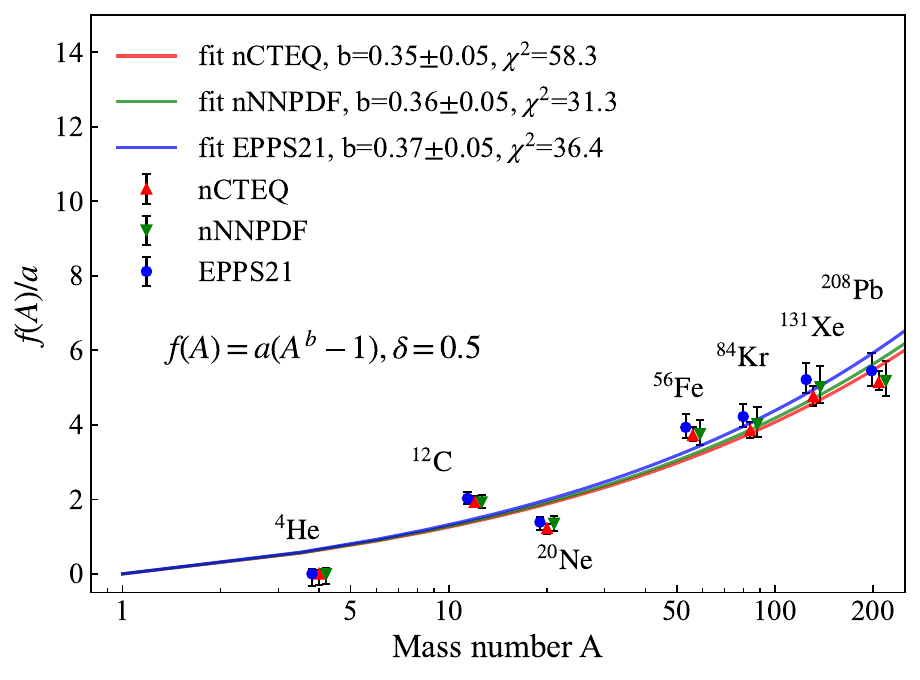}
\end{overpic}
\begin{overpic}[width = 0.49 \textwidth]{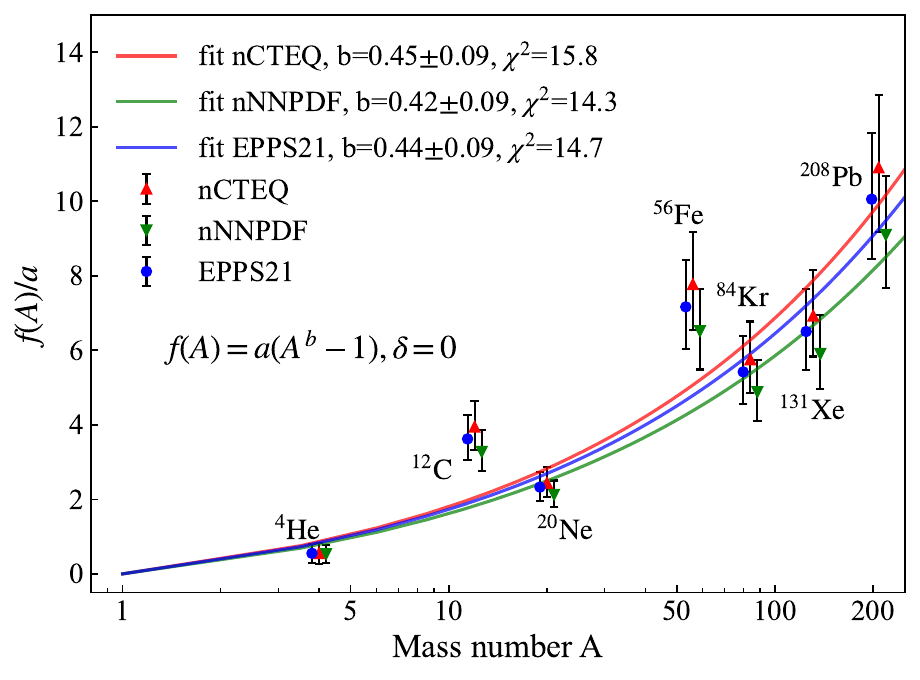}
\end{overpic}
\begin{overpic}[width = 0.49 \textwidth]{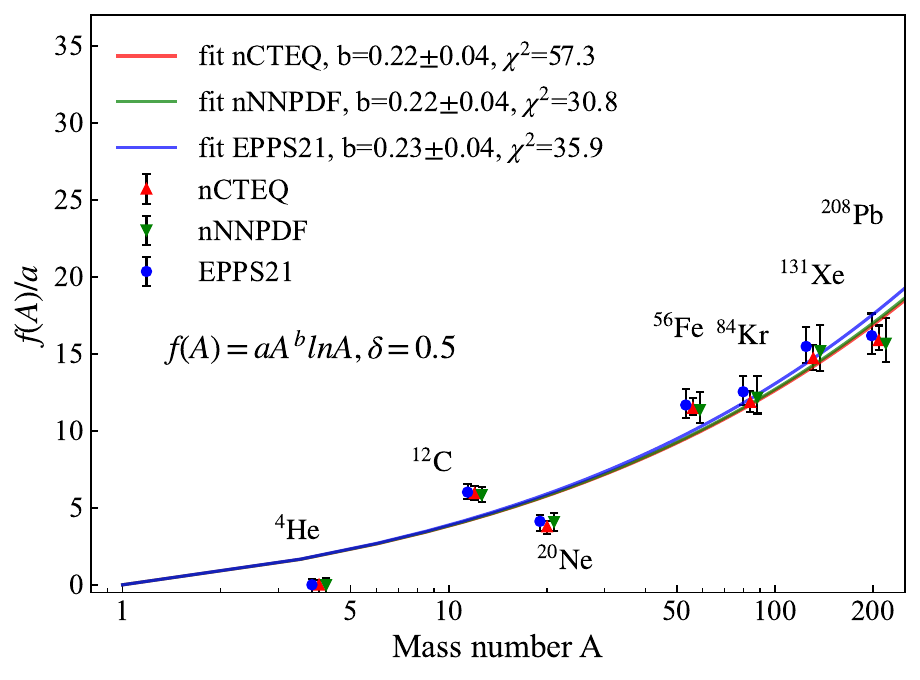}
\end{overpic}
\begin{overpic}[width = 0.49 \textwidth]{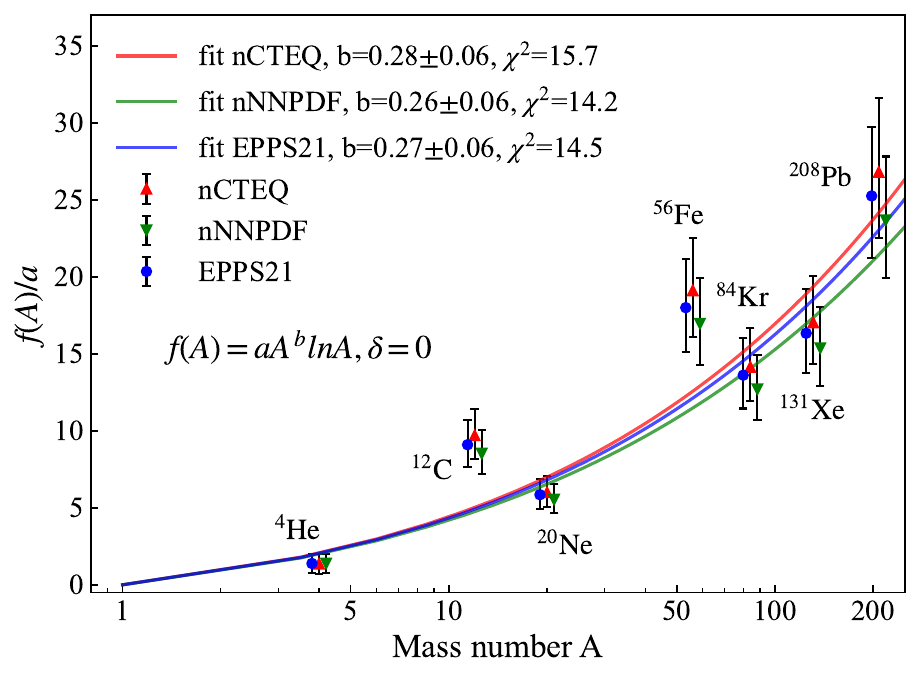}
\end{overpic}
\begin{overpic}[width = 0.49 \textwidth]{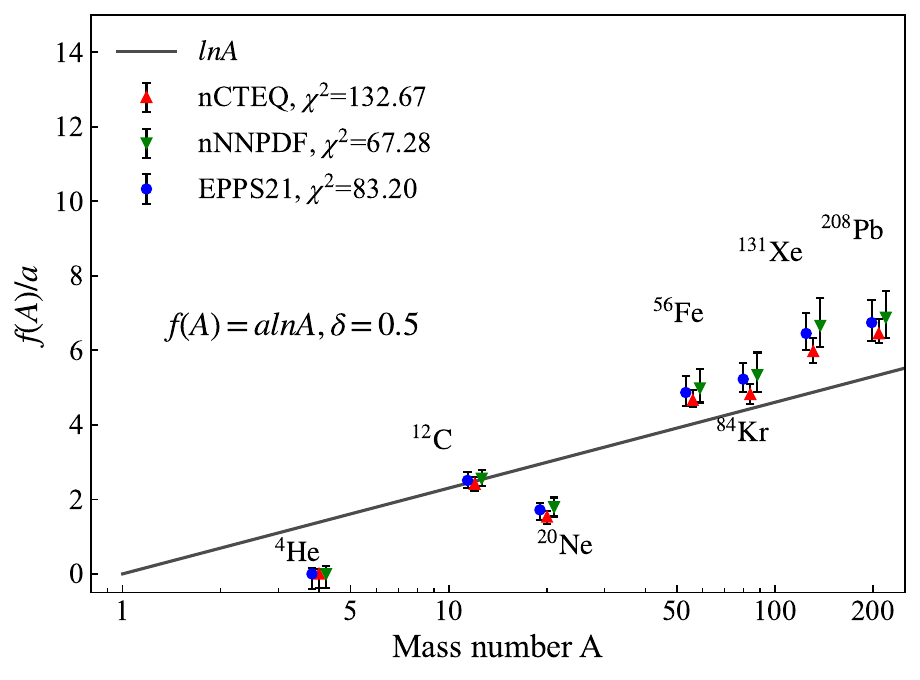}
\end{overpic}
\begin{overpic}[width = 0.49 \textwidth]{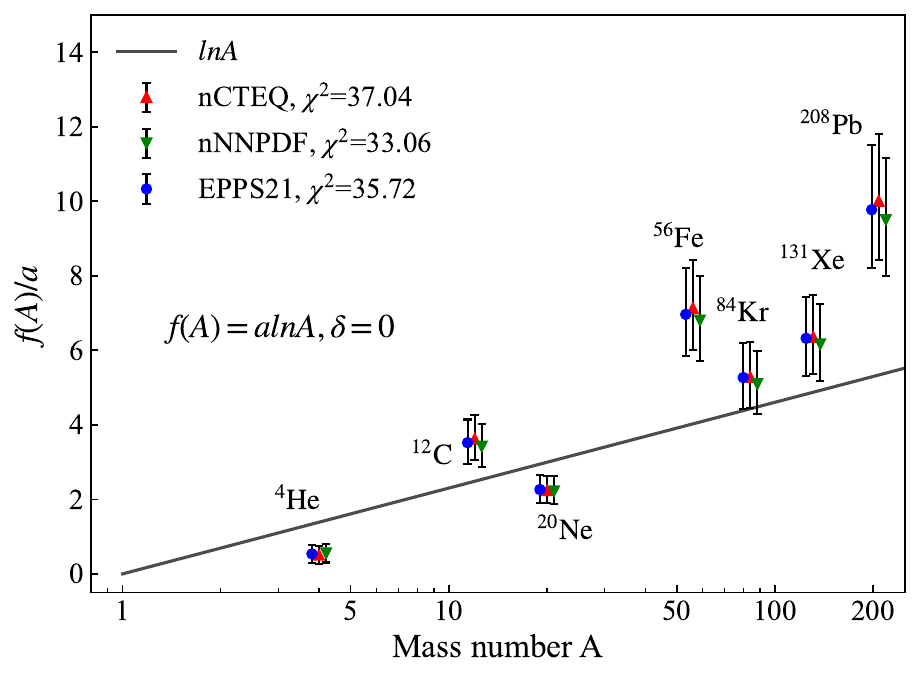}
\end{overpic}
\caption{
$A$ dependence of the function $f(A)$ for the $\delta=0.5$ and $\delta=0$ schemes. The extracted $f(A)$ points are obtained from separate nFF fits using different nPDF inputs and are compared with several functional forms normalized by the parameter $a$ to show the $A$-dependent shape.
}
\label{f_A_2}
\end{figure*}

As discussed in Section~\ref{ss.nuclear}, the normalization of nuclear modification is characterized by the parameter $a_0$ in Eq.~\eqref{eq.nffpar} and the function $f(A)$ in Eq.~\eqref{eq.fa}, where the data points of $f(A)$ are determined independently for each mass number $A$ within the framework.
To investigate this $A$-dependence, we perform independent fits to the extracted $f(A)$ data points using several parameterizations employed in studies of nPDFs and nFFs~\cite{nCTEQ:2023cpo, Wang_2001ff}:
\begin{equation}
f(A) = aA^b - a, \,\\
f(A) = aA^b \ln A, \,\\
f(A) = a \ln A,
\label{eq:parameterizations}
\end{equation}
where $a$ and $b$ are free parameters to be fitted, and all parameterizations are constructed to satisfy the physical constraint $f(A=1)=0$. 
The corresponding results for the $\delta=0.5$ scheme, normalized by the parameter $a$ to compare the $A$-dependent shape, are shown in the left panels of Fig.~\ref{f_A_2}. For comparison, the results obtained with the $\delta=0$ scheme are displayed in the right panels. The extracted $f(A)$ data points are obtained from separate nFF fits performed with the three different nPDF inputs, and are shown with error bars in different colors for comparison.

Consistent with our previous findings, the results show that the normalized $f(A)$ exhibits a similar behavior across different nPDFs, indicating that the determination of nuclear modifications is insensitive to the choice of nPDFs. 
A comparison between the $\delta=0.5$ and the $\delta=0$ schemes reveals that the latter exhibits noticeably larger uncertainties in the extracted $f(A)$ values, as reflected by the wider error bars. This leads to a reduced $\chi^2$ in the corresponding fits. Such behavior can be understood from the fact that the $\delta = 0$ scheme does not include any explicit $\nu$ dependence, thereby shifting more of the variation into the $A$ dependence and resulting in a stronger, but less constrained, $A$-dependent behavior.
For the parameterization $f_A = aA^b - a$, shown in Fig.~\ref{f_A_2}, the fit provides a good description of the extracted $f(A)$ values. The exponent $b$ is found to lie in the range $0.35$--$0.45$, with little dependence on the choice of nPDF set or $\nu$ scheme. This range is consistent with the expected scaling behavior between $A^{1/3}$ and $A^{2/3}$, as predicted by the multi-parton scattering mechanism~\cite{Wang_2001ff}.
The alternative parameterizations, $aA^b \ln A$ and $a \ln A$, shown in Fig.~\ref{f_A_2}, also provide a satisfactory description of the extracted $f(A)$ values. These forms introduce additional flexibility through logarithmic dependence while preserving the constraint $f(1)=0$.


\section{Predictions for pA Collisions}
\label{s.prediction}

In this section, we provide predictions for hadron production in $pp$ and $pA$ collisions.
Recent measurements by the ALICE collaboration~\cite{ALICE:2020atx} at the LHC provide high-precision $\gamma^{\rm iso}$-tagged fragmentation functions in $pp$ and $p$Pb collisions at a center-of-mass energy of $\sqrt{s_{\rm NN}} = 5.02\ \text{TeV}$. 
The measurement considers isolated photons in the range $|\eta| < 0.67$ and $12 < p_{\rm T}^{\gamma} < 40~\text{GeV}$, and reports the associated yield of charged hadrons within $|\eta| < 0.80$ and $0.5 < p_{\rm T}^{h} < 10~\text{GeV}$. 
The $\gamma^{\rm iso}$-tagged fragmentation functions are measured in different forms with respect to $z_{\rm T}=p^{h}_{\rm T}/p^{\gamma}_{\rm T}$.
This definition relates the transverse momentum of the produced hadron to that of the isolated photon and, owing to the momentum balance between the photon and the recoil parton, allows $z_{\rm T}$ to serve as a good approximation of the hadron energy fraction $z$ in FFs.

Furthermore, as in SIDIS where $\nu$ represents the virtual-photon energy in the target rest frame, a corresponding quantity can be introduced for isolated photon--hadron production in $p$Pb collisions. 
For the partonic subprocess $a+b\to\gamma+c$, the final-state parton $c$ fragments into the observed hadron. 
We define the analogue of $\nu$ as the energy of this fragmenting parton in the rest frame of the Pb nucleus,
\begin{equation}
\nu_c^{\rm Pb}
\equiv p_c^\mu u^{\rm Pb}_\mu
=
p_{T,c}\cosh\left(y_c-Y_{\rm Pb}\right),
\end{equation}
where $p_c^\mu$ is the four-momentum of the fragmenting parton, and $u_{\rm Pb}^\mu$ is the four-velocity of the Pb nucleus. In the last equality, $p_{T,c}$ and $y_c$ denote the transverse momentum and rapidity of the fragmenting parton, while $Y_{\rm Pb}$ is the rapidity of the Pb beam.
Unlike fixed-target SIDIS measurements, where the nucleus is at rest and $\nu$ is typically below $50~\mathrm{GeV}$, the corresponding fragmenting parton energy in $p$Pb collisions is strongly enhanced by the Lorentz boost of the colliding nucleus and can easily reach several hundred GeV or even TeV scale. 
For the ALICE kinematics considered here, a fragmenting parton with $p_{T,c}\simeq p_T^\gamma=12$--$40~\mathrm{GeV}$ produced near midrapidity corresponds to $\nu_c^{\rm Pb}\sim 20$--$70~\mathrm{TeV}$ in the Pb rest frame. As a consequence, $\gamma^{\rm iso}$-tagged fragmentation functions in $pp$ and $p$Pb collisions provide a useful and independent probe of possible $\nu$-dependent nuclear modifications in the fragmentation process.

\begin{figure}[htbp]
\centering

\begin{overpic}[width=0.48\textwidth]{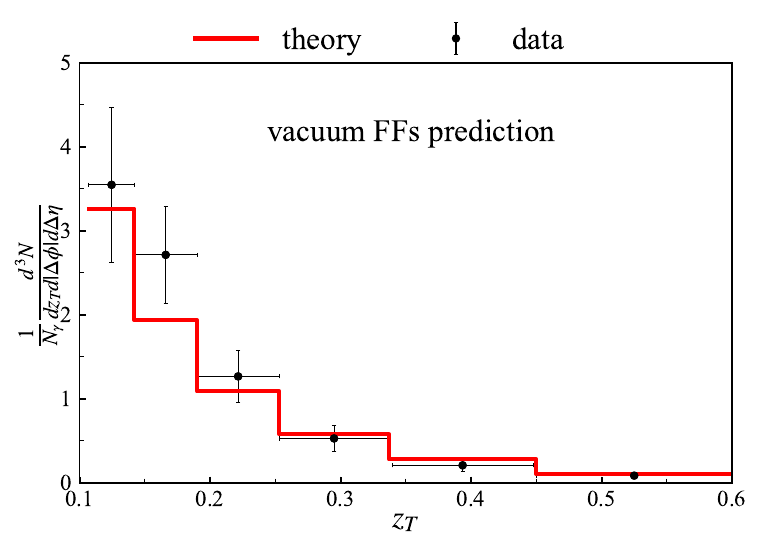}
\end{overpic}

\caption{Comparison of the $\pi^{\pm}$ NLO prediction obtained with the default vacuum FFs and ALICE measurements for $\gamma^{\rm iso}$-tagged fragmentation functions of charged hadron in $pp$ collisions at $\sqrt{s} = 5.02~\text{TeV}$.}
\label{f_prediction_pp}
\end{figure}

Figure~\ref{f_prediction_pp} shows the ALICE measurements of $\gamma^{\rm iso}$-tagged fragmentation functions in $pp$ collisions at $\sqrt{s}=5.02~\text{TeV}$, together with our NLO vacuum FFs prediction. 
It should be noted that the experimental measurements are for charged hadrons, whereas our calculation includes only the $\pi^{\pm}$ contribution. The contributions from $K^{\pm}$ and $p/\bar p$, which are expected to amount to roughly $20\%$ of the $\pi$ contribution, are not included. Therefore, the comparison should be interpreted as a $\pi$-only approximation to the charged-hadron observable. Despite this limitation, the prediction gives a reasonable description of the measured $\gamma^{\rm iso}$-tagged fragmentation functions and provides a useful baseline for the study of nuclear effects in $p$Pb collisions.

\begin{figure}[t]
\centering
\begin{overpic}[width=0.49\textwidth]{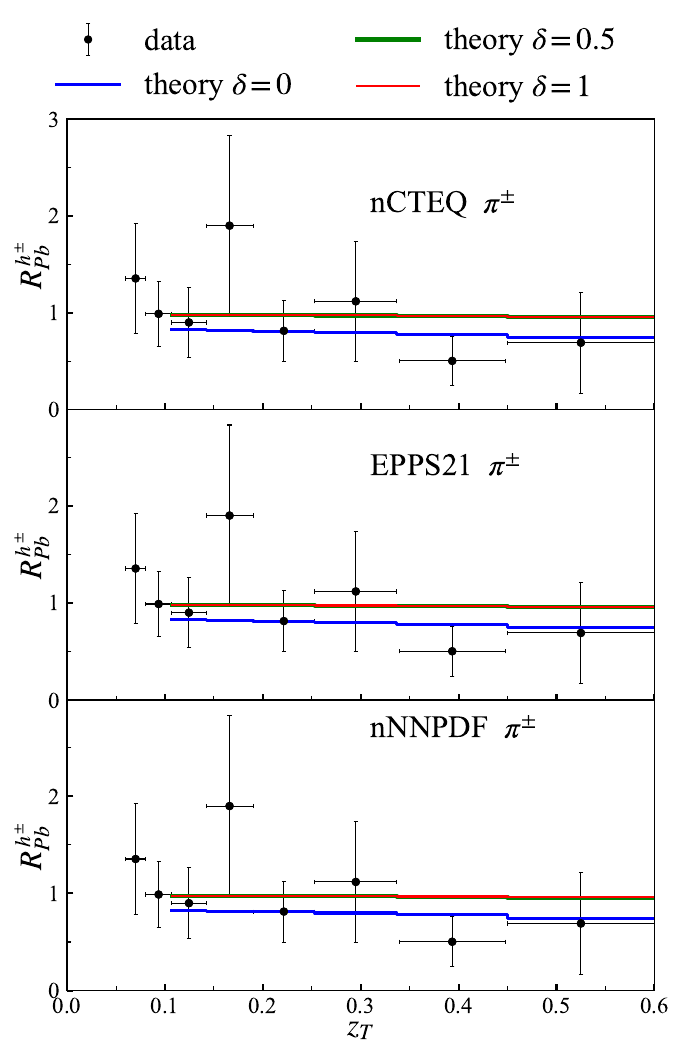}
\end{overpic}

\caption{
$p$Pb-to-$pp$ nuclear modification ratio for $\gamma^{\rm iso}$-tagged fragmentation functions at $\sqrt{s_{\rm NN}}=5.02~\mathrm{TeV}$, compared with the ALICE measurements. The three panels show predictions obtained with nFFs extracted using the nCTEQ15WZ, EPPS21, and nNNPDF nPDF inputs, respectively.
}
\label{f_prediction_pPb}
\end{figure}
Using the same $\pi$-only setup, Fig.~\ref{f_prediction_pPb} presents $p$Pb-to-$pp$ nuclear modification ratio for the $\gamma^{\rm iso}$-tagged fragmentation functions at $\sqrt{s_{\rm NN}}=5.02~\text{TeV}$. The three panels correspond to the predictions obtained with the nFFs extracted using the nCTEQ15WZ, EPPS21, and nNNPDF nPDF sets, respectively. In each case, the $\delta=0$, $0.5$, and $1$ schemes are also shown to illustrate the sensitivity to the assumed $\nu$ dependence.
For $p$Pb collisions, the theoretical predictions show a distinct pattern, where the $\delta=0.5$ and $\delta=1$ schemes are very similar over the full $z_{\rm T}$ range and remain close to unity, while the $\delta=0$ scheme tends to give smaller nuclear modification ratios.
This difference originates from the fact that the $\delta=0$ scheme does not include $\nu$ dependence in the fit, leading to a situation where the nuclear suppression is carried over to $p$A collisions, without allowing it to vary with the relevant energy scale $\nu$. As discussed above, this fragmenting parton energy scale in $p$Pb is much larger than the virtual-photon energy $\nu$ in the target rest frame. In contrast, the $\delta=0.5$ and $\delta=1$ schemes incorporate the $\nu$-dependent modification, which allows the nuclear effects to respond to the appropriate $\nu$ scale in $p$Pb collisions.
Despite these differences, all three schemes provide a reasonable description of the experimental data within the relatively large experimental uncertainties, which do not allow a clear discrimination among them. The results also show only a weak dependence on the choice of nPDF set for all $\delta$ schemes.

Overall, within the $\pi$-only setup, the present results are compatible with the $\gamma^{\rm iso}$-tagged fragmentation functions measurement in $pp$ collisions and with the corresponding $p$Pb-to-$pp$ nuclear modification
ratio. More precise measurements from LHC experiments such as ALICE, CMS, and ATLAS would be valuable for further constraining nuclear fragmentation effects and improving the discrimination among different $\delta$ schemes.


\section{Discussion and Conclusions}
\label{s.conclusion}

Understanding hadronization in the nuclear environment remains an important challenge in QCD. Collinear QCD factorization with modified nFFs offers an approach to describe it. In this framework, nFFs serve as nonperturbative inputs that effectively encode the medium-induced modifications of the hadronization process inside nuclei.
In this work, we have presented an extraction of nFFs for $\pi$, based on a comprehensive analysis of SIDIS measurements on nuclear targets. The analysis has been performed at NLO accuracy in perturbative QCD with uncertainties quantified using the Hessian method. In the kinematic region $z \sim (0.2, 0.8)$, where precise measurements are available, the extracted nFFs are well constrained. 
As for the quality of the fit, reasonable $\chi^2$ values are obtained for most datasets from various observables. With the inclusion of $\nu$-dependent nuclear measurements in the fit of $\pi$ nFFs, we are able to explore different possible forms of $\nu$-dependence of the extracted nFFs.

Our work introduces several methodological aspects in the treatment of the fit, apart from the selections of kinematics.
In previous nFF analyses, the vacuum FFs were typically fixed to existing vacuum FFs obtained mainly from global fits to high-energy experimental data, while only the nuclear modifications were fitted. However, such an approach may not fully account for the fact that SIDIS measurements on nuclear targets are predominantly performed in a relatively low $Q$ region, where the existing vacuum FFs may not provide a sufficiently accurate description. Motivated by this consideration, we adopt a framework in which vacuum FFs and nuclear modifications are extracted simultaneously, allowing the vacuum FFs to be constrained by low-energy experimental data relevant for the determination of nFFs.
To establish a reliable fitting setup for the nFFs extraction, we first performed a series of independent fits of vacuum FFs using SIDIS data on $p$ and D targets together with heavy-flavor-tagged SIA measurements. Within these studies, we investigated the stability of the extracted vacuum FFs under different theoretical setups, including variations of the perturbative accuracy between NLO and NNLO, different choices of the initial scale $Q_0 = 1.0$ and $1.3~\mathrm{GeV}$, as well as the impact of including theoretical uncertainties in the fit. Based on these studies, we selected the default setup that provides the most stable and reliable baseline for the subsequent extraction of nuclear modifications.
The vacuum FFs and nuclear modifications were then fitted simultaneously within the selected theoretical setup. The dependence of the nuclear effects on $\nu$ was first explored through different parametrization scenarios. Independent fits were performed for the $\delta=0$, $\delta=0.5$, and $\delta=1$ schemes introduced in Eq.~\eqref{eq.fa}. Comparisons among these scenarios indicate that the $\delta=1$ scheme provides a poorer description of the data, while the $\delta=0$ and $\delta=0.5$ schemes give comparable global fit qualities. In the HERMES comparison as a function of $\nu$ in fixed $z$ regions, the $\delta=0.5$ scheme gives a more balanced description of the observed increasing trend of $R_A^h$ with $\nu$.
At $Q=5~\mathrm{GeV}$ and $\nu=2~\mathrm{GeV}$, the nuclear modification of $\delta=0.5$ scheme reaches up to about $20\%$ for C and up to nearly $50\%$ for Pb, with a weaker effect observed for $d$ quarks compared to $u$ quarks.
Furthermore, the sensitivity of the extracted nFFs to the choice of initial-state nPDFs has also been examined using several nPDF sets, where the results show only a weak dependence on the nPDF input.
The nuclear-mass dependence of the extracted modifications has been investigated through independent fits of the function $f(A)$ in Eq.~\eqref{eq.fa} for different nuclear targets. The results indicate a power-like behavior with an effective exponent in the range $0.35$--$0.45$, which provides a good description of the data. This behavior is consistent with expectations from multiple parton scattering mechanisms.
As an application, the extracted nFFs have been used to provide predictions for $\gamma^{\rm iso}$-tagged fragmentation functions in $pp$ and $p$Pb collisions at LHC energies. Within the $\pi$-only approximation used in this work, the comparison with ALICE data shows a reasonable agreement within relatively large uncertainties. This suggests that such observables can provide useful complementary constraints on nuclear modifications in high-energy $pA$ collisions. More precise measurements from LHC experiments would be highly valuable for further constraining nuclear fragmentation effects and improving the sensitivity to different implementations of the $\nu$ dependence.

This work is a continuation of the previous NPC FFs analysis with extension to nFFs. 
Taking into account the new constraints provided by SIDIS data with $\nu$ information, the present analysis establishes a framework for future studies of nFFs and hadronization in nuclear environments. The extracted FFs are publicly available in LHAPDF format.

\noindent
\begin{acknowledgments}

The work of J.G. is supported by the National Natural Science Foundation of China (NSFC) under Grant No. 12275173, Shanghai Municipal Education Commission under Grant No. 2024AIZD007, and open fund of Key Laboratory of Atomic and Subatomic Structure and Quantum Control (Ministry of Education).
H.X. is supported by the NSFC under Grant Nos. 12525508, 12475139.
Y. Zhao is supported by the NSFC under Grant No. U2032105 and the CAS Project for Young Scientists in Basic Research No. YSBR-117.
\end{acknowledgments}

%


\bibliography{refs}
\vspace*{10pt}
\appendix
\part*{Appendices}
\addcontentsline{toc}{part}{Appendices}
\section{Impact of nPDFs}
\label{appendix}

We conducted a benchmark study to quantify the impact of different nPDF sets on the theoretical predictions. This step is important in order to disentangle possible uncertainties associated with the initial-state nuclear structure from those originating from the final-state fragmentation process. Since both nPDFs and nFFs enter the SIDIS cross sections simultaneously within the collinear factorization framework, it is necessary to verify that the extracted nuclear modifications are not biased by the particular choice of nPDF parametrization.
In this benchmark, the baseline nuclear fragmentation functions were first extracted using the nCTEQ15WZ nPDF set. The extracted nFFs were then kept fixed, while the SIDIS cross sections were recalculated using alternative nPDF sets, namely EPPS21 and nNNPDF. In this way, the role of the initial-state nuclear input can be isolated, allowing a direct assessment of how the nPDF choice propagates into the final observable predictions.

\begin{figure*}[htbp]
\centering
\begin{overpic}[width = 0.9 \textwidth]{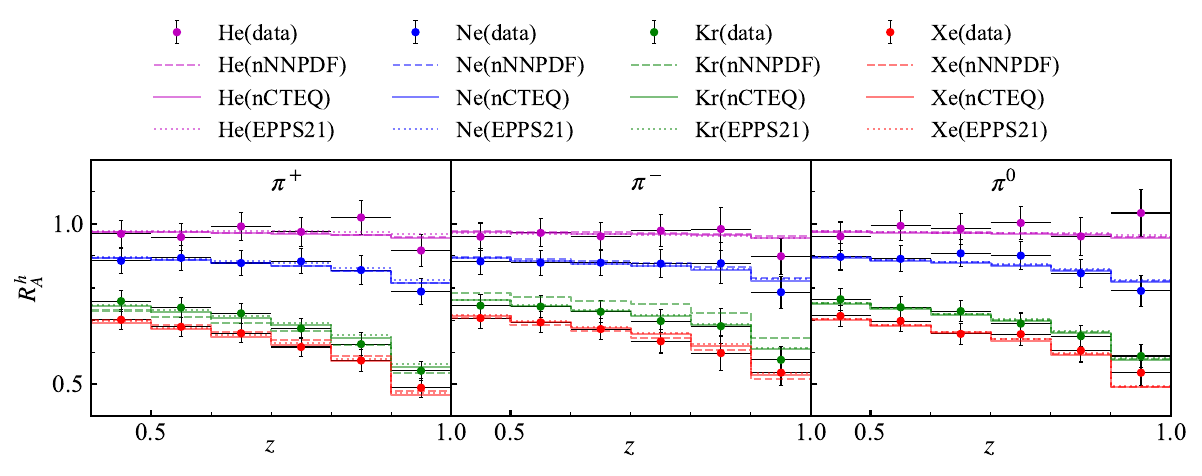}
\end{overpic}
\caption{
Benchmark comparison of measured and predicted nuclear modifications for
pion production in HERMES experiments~\cite{HERMES:2007plz}
using different nPDF inputs while keeping the extracted nFFs fixed.
}
\label{f.nff_nFF_inf_inv0_hermes}
\end{figure*}

\begin{figure*}[htbp]
\centering
\begin{overpic}[width = 0.65 \textwidth]{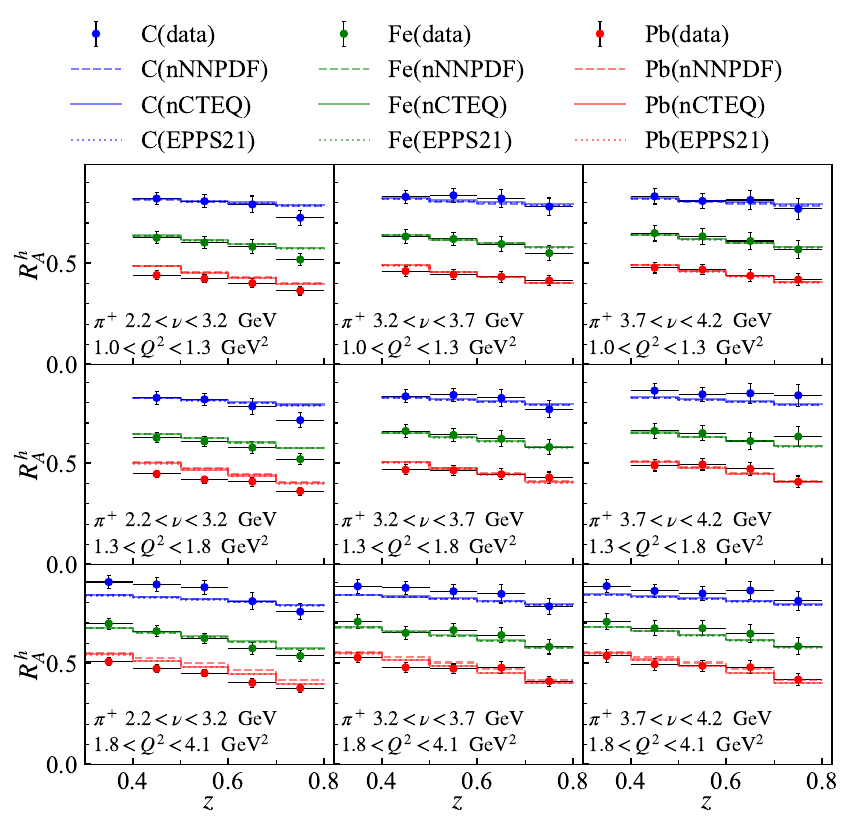}
\end{overpic}
\caption{
Similar to Fig.~\ref{f.nff_nFF_inf_inv0_hermes}, but for the CLAS measurements~\cite{CLAS:2021jhm} shown as functions of $z$ and $\nu$.
}
\label{f.nff_nFF_inf_inv0_clas_pi}
\end{figure*}

For clarity of presentation, representative HERMES and CLAS datasets are shown in Figs.~\ref{f.nff_nFF_inf_inv0_hermes} and \ref{f.nff_nFF_inf_inv0_clas_pi}. As can be seen from these figures, replacing the nPDF input from nCTEQ15WZ to EPPS21 or nNNPDF while keeping the nFFs fixed leads only to very small changes in the resulting predictions. The overall agreement with the experimental data remains essentially unchanged for all considered nPDF sets.
Overall, these results further prove that the extracted nFFs exhibit good stability against variations of the nPDF input. The residual dependence on the choice of nPDF set is significantly smaller than the differences associated with the various $\delta$-scaling scenarios considered in the main analysis.

\end{document}